\documentclass[preprint,sort&compress,12pt]{elsarticle}

\usepackage{graphicx}
\usepackage{amsmath}

\usepackage{natbib}
\usepackage{amssymb}

\usepackage{amsthm}

\usepackage{lineno}
\usepackage{subfig}
\usepackage{enumerate}
\usepackage{fullpage}
\usepackage{algorithm}
\usepackage{algpseudocode}
\usepackage{xcolor}
\usepackage[colorinlistoftodos]{todonotes}

\biboptions{sort,compress} 

\usepackage{xcolor}

\newcommand*\patchAmsMathEnvironmentForLineno[1]{%
  \expandafter\let\csname old#1\expandafter\endcsname\csname #1\endcsname
  \expandafter\let\csname oldend#1\expandafter\endcsname\csname end#1\endcsname
  \renewenvironment{#1}%
     {\linenomath\csname old#1\endcsname}%
     {\csname oldend#1\endcsname\endlinenomath}}%
\newcommand*\patchBothAmsMathEnvironmentsForLineno[1]{%
  \patchAmsMathEnvironmentForLineno{#1}%
  \patchAmsMathEnvironmentForLineno{#1*}}%
\AtBeginDocument{%
\patchBothAmsMathEnvironmentsForLineno{equation}%
\patchBothAmsMathEnvironmentsForLineno{align}%
\patchBothAmsMathEnvironmentsForLineno{flalign}%
\patchBothAmsMathEnvironmentsForLineno{alignat}%
\patchBothAmsMathEnvironmentsForLineno{gather}%
\patchBothAmsMathEnvironmentsForLineno{multline}%
}

\usepackage{graphicx}
\usepackage{amssymb}
\usepackage{amsthm}
\usepackage{bbm}
\usepackage{bm}
\usepackage{lineno}
\usepackage{fullpage}
\usepackage[colorinlistoftodos]{todonotes}
\usepackage{url}
\usepackage{listings}
\usepackage[colorlinks=true]{hyperref}

\usepackage{color,soul}
\usepackage{mathtools}

\definecolor{lightblue}{rgb}{.90,.95,1}
\definecolor{darkgreen}{rgb}{0,.5,0.5}

\definecolor{lightgreen}{rgb}{.90,1,0.90}

\newcommand{\bs}[1]{\boldsymbol{#1}}

\usepackage{gensymb}
\usepackage{array}
\usepackage{changes}
\usepackage{enumerate}
\newcolumntype{P}[1]{>{\centering\arraybackslash}m{#1}}

\newcolumntype{L}[1]{>{\raggedright\let\newline\\\arraybackslash\hspace{0pt}}m{#1}}
\newcolumntype{C}[1]{>{\centering\let\newline\\\arraybackslash\hspace{0pt}}m{#1}}
\newcolumntype{R}[1]{>{\raggedleft\let\newline\\\arraybackslash\hspace{0pt}}m{#1}}
\usepackage{todonotes} 

\graphicspath{ {./figs/} }

\journal{}

\begin{document}

\begin{frontmatter}





\title{Representation of Reynolds Stress Perturbations with Application in 
Machine-Learning-Assisted Turbulence Modeling}





\author[vt]{Jinlong Wu}
\author[vt]{Rui Sun}
\author[icl]{Sylvain Laizet}
\author[vt]{Heng Xiao\corref{corxh}}
\ead{hengxiao@vt.edu}
\cortext[corxh]{Corresponding author. Tel: +1 540 231 0926}
\address[vt]{Department of Aerospace and Ocean Engineering, Virginia Tech, Blacksburg, VA 24060, United States}
\address[icl]{Department of Aeronautics, Imperial College London, London, United Kingdom}

\begin{abstract}
  Numerical simulations based on Reynolds-Averaged Navier--Stokes (RANS) equations are widely used in engineering design and analysis involving turbulent flows. However, RANS simulations are known to be unreliable in many flows of engineering relevance, which is largely caused by model-form uncertainties associated with the Reynolds stresses. Recently, a machine-learning approach has been proposed to assist RANS modeling by building a functional mapping from mean flow features to discrepancies in RANS modeled Reynolds stresses as compared to high-fidelity data. However, it remains a challenge to represent discrepancies in the Reynolds stress eigenvectors in machine learning due to the requirements of spatial smoothness, frame-independence, and realizability. In this work,  we propose three schemes for representing perturbations to the eigenvectors of RANS modeled Reynolds stresses: (1) discrepancy-based Euler angles, (2) direct-rotation-based Euler angles, and (3) unit quaternions.  We compare these metrics by performing \textit{a priori} and \textit{a posteriori} tests on two canonical flows: fully developed turbulent flows in a square duct and massively separated flows over periodic hills. The results demonstrate that the direct-rotation-based Euler angles representation lacks spatial smoothness while the discrepancy-based Euler angles representation lacks frame-independence, making them unsuitable for being used in machine-learning-assisted turbulence modeling. In contrast, the representation based on unit quaternion satisfies all the requirements stated above, and thus it is an ideal choice in representing the perturbations associated with the eigenvectors of Reynolds stress tensors. This finding has clear importance for uncertainty quantification and machine learning in turbulence modeling and for data-driven computational mechanics in general.
\end{abstract}

\end{frontmatter}



\section*{Notation}
We summarize the notations to illustrate the convention of the nomenclature used in this
paper.  Upper case letters (e.g., $\mathbf{Q}$) indicate matrices or tensors; lower case letters with bold font (e.g., $\mathbf{n}$) indicate vectors; undecorated letters in lower cases indicate scalars. Tensors (matrices) and vectors are also
  indicated with index notations, e.g., $R_{ij}$ and $u_i$ with $i, j = 1, 2, 3$. In this paper, $i$ and $j$ are used with tensor indices while $\alpha$ is used as general indexes. The ensemble average is indicated by $\overline{\Box}$. The superscript $\Box^o$ denotes original quantities given by the Reynolds-Averaged Navier--Stokes simulations (baseline for the perturbations), and the superscript $\Box^*$ represents the truth or the target of perturbations. A list of  nomenclature is presented in Appendix.

\section{Introduction}
\label{sec:intro}

\subsection{Challenges and new developments in turbulence modeling}
Numerical simulations based on Reynolds-Averaged Navier--Stokes (RANS) equations are still the
dominant tool for industry applications involving turbulent flows, even though the rapid growth of
computational resource has greatly expanded the reach of the high-fidelity simulation methods such
as Direct Numerical Simulation (DNS) and Large Eddy Simulation (LES) in the past few
decades. However, commonly used RANS models (e.g., $k$-$\varepsilon$ models, $k$-$\omega$ models and
S-A models) are known to be unreliable in many flows such as those with three-dimensional
separations, strong pressure gradients or mean streamline
curvature~\cite{wilcox1998turbulence}. This lack of accuracy mainly originates from a large
discrepancy between the modeled and the true Reynolds stresses, leading to the unreliable
predictions of other quantities of interests (QoIs) such as mean velocity, mean pressure, surface
friction, and drag and lift forces.

In light of the decades-long stagnation in traditional RANS modeling, several researchers developed
machine-learning-assisted turbulence
models~\cite{wang2017physics,ling2016reynolds,singh2016machine}. These efforts aimed at leveraging
machine learning algorithms and large amounts of data made available by advances in experimental
techniques and computational sciences. Duraisamy and
co-workers~\cite{tracey2015machine,singh2016machine} identified the target of
machine-learning-assisted models as the multiplicative discrepancy in the production term of the
transport equations of turbulent quantities (e.g., $\tilde{\nu}_t$ in the S--A model). Although
recent studies showed the potential of extrapolation capabilities of this approach among different
flows~\citep{singh16using,singh2016machine}, such capabilities are potentially limited by the lack
of physical anchoring and uniqueness of the multiplicative discrepancy term. On the other hand, Ling et
al.~\cite{ling2016reynolds} incorporated physical knowledge in designing the architecture of a
machine-learning-assisted model by adopting an invariant-set-based representation of Reynolds stress
anisotropy tensor~\cite{popebook}.  Their approach has clear physical justification, but note that
their aim was to replace the traditional turbulence models with a data-driven, machine-learning-based
counterpart.  Xiao and co-workers~\cite{xiao2017perspectives,wang2017physics} argued that
data-driven models should be used to assist and complement, rather than replace, the traditional
turbulence models. They justified their philosophical argument with the fact that currently used
turbulence models condensed a lot of physical and theoretical insights, which in turn were distilled
from large amounts of data and empirical knowledge in decades of engineering practice. In addition,
these models have achieved great successes in engineering turbulent flow simulations despite the
above-mentioned limitations. As such, Wang et al.~\cite{wang2017physics} proposed a machine-learning-assisted turbulence modeling framework, where they identified the perturbations that can correct RANS-modeled Reynolds stresses to the true Reynolds stress as the target of learning.  While this work focuses on RANS modeling, we note that several data-driven approaches have also been proposed in the context of LES to improve sub-grid scale stress models~\cite{king2016autonomic} and scalar fluxes models~\cite{king2016autonomic,vollant2017subgrid}.

\subsection{Stress perturbations in machine-learning-assisted RANS modeling}

Wang et al.~\cite{wang2017physics} represented the perturbations to RANS modeled Reynolds stress by
following the decomposition method of Iaccarino and co-workers in their model-form uncertainty
quantification work~\cite{emory2013modeling,emory2011modeling,emory14estimate}.  Iaccarino et
al.~\cite{emory2013modeling} decomposed the Reynolds stress anisotropic tensor into eigenvalues and eigenvectors, and
then they used Barycentric triangle to provide a realizability map for the eigenvalues.  The
Barycentric triangle is equivalent to the well-known Lumley
triangle~\cite{banerjee2007presentation,lumley1977return}.  These two equivalent maps provide a
convenient way to ensure the realizability of the perturbed Reynolds stresses by bounding the mapped
eigenvalues within the respective triangles. The theoretical foundation is that, after the
corresponding transformations~\cite{banerjee2007presentation}, a Reynolds stress tensor must reside within
or on the edge of the Lumley or Barycentric triangle.  However, much less work has been devoted to
representing the perturbations to the eigenvectors of the RANS modeled Reynolds stress tensor.

Perturbations to eigenvectors are much more difficult to impose compared with that to the
eigenvalues, which is due to several reasons. First, there is no straightforward bound on
eigenvector perturbations -- Reynolds stress realizability as represented by the Barycentric
triangle only provides a bound for the eigenvalues and not for the eigenvectors.  In the context of
uncertainty estimation, researchers investigated several representations of eigenvector
perturbations. Thompson et al.~\cite{roney16strategy} used the Reynolds stress transport equations to
constrain the eigenvector perturbations based on the bounds on the eigenvalue perturbations. Their
work can potentially address the challenge of bounding eigenvector perturbations. However, it should be noted that the Reynolds stress transport equations have several unclosed source terms (e.g., pressure-strain tensor, triple correlation) that must be estimated,
which makes the transport-equation-based constraint a soft rather than a strict one.  On the other hand,
Iaccarino et al.~\cite{iaccarino2017eigenspace} proposed an approach to augment their eigenvalue
perturbations by using the maximum and minimum of turbulence production as bounds for perturbing the
eigenvectors.  Such bounds are physically sound, albeit not necessarily as mathematically rigorous as
that in~\cite{roney16strategy}.  In addition to the lack of straightforward bounds, another
challenge is that any perturbations introduced to the eigenvectors must retain their
orthonormality. This is to ensure the symmetry and realizability of the perturbed Reynolds stress
tensor. Such a requirement immediately rules out the option of introducing componentwise discrepancy
tensor to eigenvectors or to the Reynolds stress tensor itself. Instead, a straightforward method to retain
such orthonormality is to represent the eigenvector perturbations as a three-dimensional rigid-body
rotation.  In this spirit, Wang et al.~\cite{mfu5} introduced such perturbations to the eigenvectors
by using Euler angles for quantifying RANS model-form uncertainties~\cite{mfu5}.


%

The work reviewed above on introducing perturbations to modeled Reynolds stresses are all concerned
with RANS model-form uncertainty estimations. A closely related topic is machine-learning-assisted
turbulence modeling as in the framework of~\cite{wang2017physics}, where the objective is to predict
the perturbations $\Delta \mathbf{R}$ needed to correct the modeled Reynold stresses
$\mathbf{R}^{rans}$ to the true $\mathbf{R}^{*}$.  They used machine learning to train a function
$f:\, \mathbf{q} \mapsto \Delta \mathbf{R}$ between mean flow features $\mathbf{q}$ and Reynolds stress
perturbations $\Delta \mathbf{R}$. By correcting the RANS-predicted Reynolds stresses towards the DNS counterparts, such
perturbations are expected to improve the predictions of the velocity field and other quantities of
interest. Wang et al.~\cite{wang2017physics} used the random forest to learn such perturbations from a
database of training flows with DNS data. Wu et al.~\cite{wu2017priori} further compared several metrics for the assessment of the prediction confidence of the machine-learning-assisted turbulence modeling. In the context of machine-learning-assisted RANS modeling,
the bounds for the eigenvectors perturbations are not required explicitly; instead, the main concern
is to ensure the orthonormality of the machine-learning-corrected Reynolds stress eigenvectors. To
this end, they used Euler angles to represent the eigenvectors perturbations as rigid-body rotations
following the work by Wang et al.~\cite{mfu5}.  In fact, the representation of rigid-body rotation has attracted much more
attention in robotics and computer vision~\cite{huynh2009metrics}, where different approaches
including those based on Euler angles~\cite{kuffner2004effective}, unit
quaternion~\cite{horn1987closed} and rotation matrices of direction-cosines~\cite{heeger1990simple}
have been evaluated and compared.

However, the usages of the above approaches to representing the eigenvectors perturbations to Reynolds stress
tensors pose unique requirements in the context of machine-learning-assisted modeling.  In
particular, the functional form $f:\, \mathbf{q} \mapsto \Delta \mathbf{R}$ between mean flow
features and desired Reynolds stress perturbations should be smooth to ensure that $f$ can be
learned from data~\cite{domingos2012few}. Otherwise, the machine learning algorithms tend to fit the noises rather than the true functional relation.  Another requirement is the frame-independence of the representation of Reynolds stress perturbations. In this work, we first address the smoothness requirement by comparing three representations of eigenvectors perturbations via $\textit{a priori}$ tests in Sec.~\ref{sec:a-priori}. We then evaluate the performances of Euler angles and unit quaternion in the context
of the machine-learning-assisted turbulence modeling in Sec.~\ref{sec:a-posteriori}.


\subsection{Novelty and potential impact of present work}
The novelty of the present contribution is twofold.  First, we explored several alternatives in
representing the perturbations on Reynolds stress eigenvectors as rigid-body rotations, which
ensure the orthonormality of the perturbed eigenvectors by construction and thus the symmetry and positive definiteness
(realizability) of the perturbed Reynolds stresses. Second, we performed a comprehensive comparison
of two types of representation of rigid-body rotations, Euler angles and unit quaternions, in light of
the two requirements posed by machine-learning-assisted turbulence modeling, i.e., smoothness and
frame-independence. The assessment demonstrates that the unit-quaternion-based representation
satisfies both requirements, making it a superior to Euler-angle-based representations of
eigenvectors perturbation for machine-learning-assisted turbulence modeling.  This finding has clear
importance in data-driven turbulence modeling and data-driven computational mechanics in
general. Moreover, it also has implications in other fields such as plasticity, where sequential
increments of stress tensors are used to find a path from the current stress state to the new state.

The rest of the paper is organized as follows. Section~\ref{sec:PIML} summarizes the
machine-learning-assisted turbulence modeling framework of Wang et
al.~\cite{wang2017physics}. Section~\ref{sec:metric} introduces three representations of the
eigenvectors perturbations to the Reynolds stress tensor, including direct-rotation-based Euler
angles, discrepancy-based Euler angles, and the unit quaternion. Section~\ref{sec:results} presents
the results in evaluating these three representations of eigenvectors perturbations. Finally,
Section~\ref{sec:conclusion} presents the conclusions.

\section{Summary of machine-learning-assisted turbulence modeling framework}
\label{sec:PIML}

\subsection{Origin of RANS model-form uncertainty}

The Navier--Stokes (NS) equations for incompressible flows with a constant density $\rho$ can be
written as follows:
\begin{subequations} \label{eq:ns}
  \begin{align}
    \frac{\partial u_i}{\partial t}+\frac{\partial \left( u_i u_j \right)}{\partial x_j} = &
    -\frac{1}{\rho}\frac{\partial {p}}{\partial x_i} +\nu\frac{\partial^2 u_i}{\partial x_j \partial
      x_j}     \label{eq:ns-momentum}  \\
    \textrm{and} \quad \frac{\partial u_i}{\partial x_i} = 0 \label{eq:ns-mass}
\end{align}
\end{subequations}
where $u_i$ and $p$ are instantaneous velocity and pressure, respectively; $t$ and $x_i$ are time
and space coordinates, respectively; $\nu$ is the kinematic viscosity.  Solving the Navier--Stokes
equations directly would necessitate resolving a wide range of spatial and temporal scale, which would
incur prohibitive computational costs. Therefore, when simulating turbulent flows in
engineering, the instantaneous fields $u_i$ and $p$ in the NS equations are usually decomposed into
their means ($\overline{u}_i$ and $\overline{p}$, respectively) and the fluctuations ($u'_i$ and $p'$) around the means, i.e.,
\begin{align}
  u_i & =   \overline{u}_i + u'_i \\
  p & = \overline{p} + p'
\end{align}
Substituting the Reynolds decomposition above into the NS equation yields the RANS equations, which
describe the mean velocities and pressure:
\begin{subequations} \label{eq:ns}
  \begin{align}
    \frac{\partial \overline{u}_i}{\partial t}+\frac{\partial \left( \overline{u}_i
        \overline{u}_j \right)}{\partial x_j} = & -\frac{1}{\rho}\frac{\partial
      {\overline{p}}}{\partial x_i} +\nu\frac{\partial^2 \overline{u}_i}{\partial x_j \partial
      x_j} - \frac{\partial \langle {u}_i' {u}_j' \rangle}{\partial
      x_j}    \label{eq:ns-momentum}  \\
    \textrm{and} \quad \frac{\partial \overline{u}_i}{\partial x_i} = 0 \label{eq:ns-mass}
\end{align}
\end{subequations}
where negative of the velocity fluctuation covariance $\langle u_i' u_j' \rangle$ is referred to as
Reynolds stress and is denoted as $R_{ij}$ or $\mathbf{R}$ for simplicity. This term needs to be
modeled. Note that the Reynolds stress $\mathbf{R}$ is a positive semidefinite tensor. It is a
consensus that in incompressible, fully turbulent flows (i.e., flows without transition, heat
transfer, or compressible effects), the modeled Reynolds stress term is the main source of
model-form uncertainties in RANS simulations~\cite{oliver2011bayesian,popebook}. It is thus natural
to focus on this term when estimating, inferring, or reducing RANS model uncertainties. At any point
in the flow field, the true Reynolds stress $\mathbf{R}^{*}$ can be written as the sum of the RANS-modeled value and a
discrepancy term, i.e., $\mathbf{R}^{*} = \mathbf{R}^{rans} + \Delta \mathbf{R}$.  
  Aiming at estimating the Reynolds stress discrepancy $\Delta \mathbf{R}$, Wang et
  al.~\cite{wang2017physics} proposed a machine-learning-assisted turbulence modeling framework,
  which is detailed below.

\subsection{Machine-learning-assisted turbulence modeling framework}
Machine learning is in fact a concept familiar to most physical scientists, although they are predominantly used in commercial applications currently. Machine learning involves three steps: (1) postulate a model, which maps input to output and is controlled by a set of adjustable model parameters; (2) fit  (i.e., learn) the parameters to given training data; and (3) use the fitted model to predict for unseen inputs.

The essence of the machine-learning-assisted turbulence modeling framework of Wang et al.~\cite{wang2017physics} is
to predict the discrepancy term $\Delta \mathbf{R}$ by learning a model from high-fidelity simulation (e.g, DNS) data. This is achieved by learning a functional mapping $f: \mathbf{q} \mapsto \Delta \mathbf{R} $,
where $\mathbf{q}$ indicates mean flow features obtained from RANS simulations, e.g., mean pressure
gradient, mean flow curvature, all normalized with local quantities.  In machine learning
terminology the discrepancy $\Delta \mathbf{R}$ is referred to as \emph{responses}, the feature
vector $\mathbf{q}$ as the \emph{inputs}, and the mappings $f$ as \emph{regression functions}. The
flows used to train the \emph{regression functions} is referred to as \emph{training flows}, and the
flow to be predicted as \emph{test flow}.

Wang et al.~\cite{wang2017physics} used a group of hand-crafted ten features $q_{i}$ based on the
RANS simulated mean flow fields (velocity $\overline{u}_i$ and pressure $\overline{p}$) as inputs
$\mathbf{q}$ to the regression function. In a more recent work~\cite{wang2017predictive}, an
additional 47 mean flow features were constructed as the invariant set $\{ \mathbf{S},
\mathbf{\Omega}, \nabla p, \nabla k \}$, where $\mathbf{S}$ denotes the strain-rate tensor,
$\bs{\Omega}$ the rotation-rate tensor, $\nabla p$ the pressure gradient and $\nabla k$ the gradient
of turbulent kinetic energy, all of which are obtained from RANS-simulated mean flow fields.

Wang et al.~\cite{wang2017physics} used a
physics-based perturbation via the following
decomposition to represent the Reynolds stress perturbations
 $\Delta \mathbf{R}$~\cite{emory2013modeling,banerjee2007presentation}:
\begin{equation}
  \label{eq:tau-decomp}
  \mathbf{R} = 2 k \left(  \frac{\mathbf{I}}{3} +  \mathbf{A} \right)
  = 2 k \left( \frac{\mathbf{I}}{3} + \mathbf{V} \mathbf{\Lambda} \mathbf{V}^T \right) 
\end{equation}
where $k$ is the turbulent kinetic energy, which indicates the magnitude of $\mathbf{R}$;
$\mathbf{I}$ is the second order identity tensor; $\mathbf{A}$ is the anisotropy tensor; $\mathbf{\Lambda} = \textrm{diag}[\lambda_1, \lambda_2,
\lambda_3]$ and $\mathbf{V} = [\mathbf{v}_1, \mathbf{v}_2, \mathbf{v}_3]$ are the eigenvalues and orthonormal eigenvectors of $\mathbf{A}$, representing its shape
(i.e., aspect ratio) and orientation, respectively.  In summary, this decomposition maps the Reynolds
stress to $(k, \mathbf{\Lambda}, \mathbf{V}$). The eigenvalues matrix $\mathbf{\Lambda}$ have two degrees of freedom due to the constraint
$\lambda_1+\lambda_2+\lambda_3=0$, and they can be mapped to a Barycentric coordinates,
where the Reynolds stress realizability can be imposed. The eigenvectors $\mathbf{V}$ consist of
three orthonormal vectors (nine elements in total) but has only three degrees of freedom.

\begin{figure}[!htbp]
  \centering
  \includegraphics[width=0.44\textwidth]{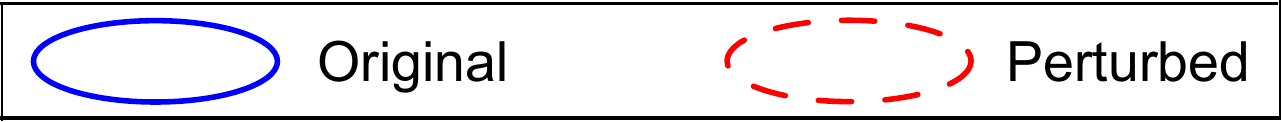}\\
  \subfloat[TKE $k$]{\includegraphics[width=0.3\textwidth]{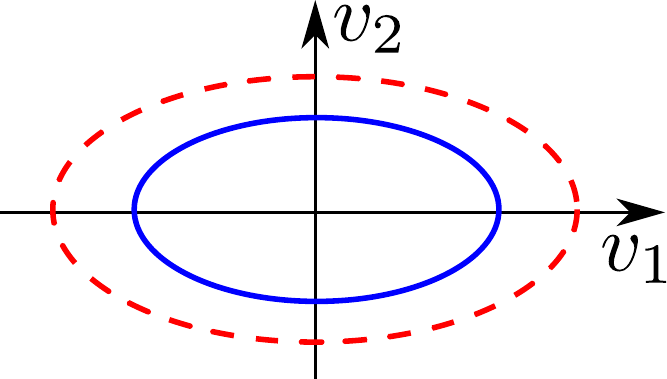}}\
  \subfloat[Eigenvalues $\mathbf{\Lambda}$]{\includegraphics[width=0.3\textwidth]{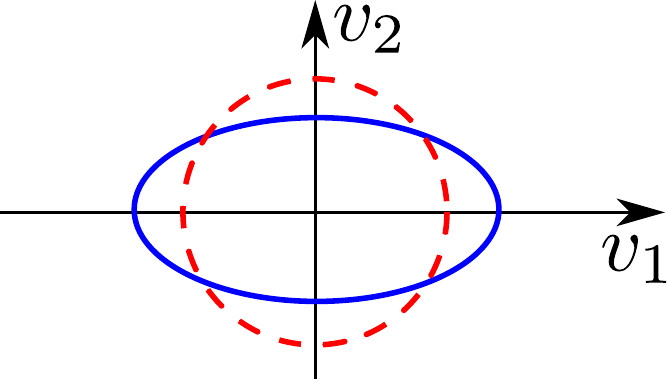}}\
  \subfloat[Eigenvectors $\mathbf{V}$]{\includegraphics[width=0.3\textwidth]{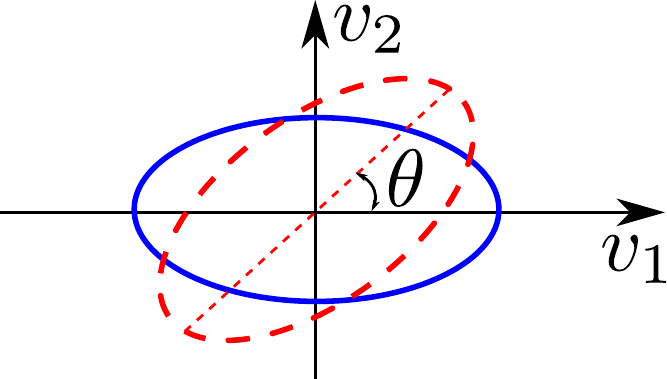}}\\
  \caption{The perturbations on (a) the magnitude (turbulent kinetic energy $k$), (b) eigenvalues $\mathbf{\Lambda}$, and (c) eigenvectors $\mathbf{V}$  of a Reynolds stress tensor $\mathbf{R}$. For clarity, the tensor is represented in the two-dimensional space as an ellipsis instead of a three-dimensional ellipsoid. In addition, perturbations on $k$, $\boldsymbol{\Lambda}$, and $\mathbf{V}$ are shown individually on each panel and not as three consecutive perturbations.}
\label{fig:perturb-scheme}
\end{figure}

Therefore, if we visualize the Reynolds stress as an ellipsoid~\cite{simonsen2005turbulent}, the
discrepancy between the modeled Reynolds stress $\mathbf{R}$ and the corresponding true $\mathbf{R}^*$
can be formulated as three consecutive transformations (i.e., perturbations):
\begin{enumerate}
\item The size of the ellipsoid (magnitude of the tensor) is scaled by a positive factor of $\gamma_k =
  k^*/k^{rans}$ while keeping the shape the same, as illustrated in Fig.~\ref{fig:perturb-scheme}a.
\item The aspect ratio of ellipsoid is perturbed while keeping the size (sum of the three axes) and
  orientation unchanged. This is achieved by perturbing the eigenvalues, i.e., $\mathbf{\Lambda}^*
  = \mathbf{\Lambda} + \Delta \mathbf{\Lambda}$, as illustrated in Fig.~\ref{fig:perturb-scheme}b.
\item Finally, the ellipsoid experiences a rigid-body rotation, which is represented as $\mathbf{V}^* = \mathbf{Q} \mathbf{V}$,
  where $\mathbf{Q}$ is a rotation matrix. This perturbation is illustrated in Fig.~\ref{fig:perturb-scheme}c.
\end{enumerate}

The objective of the machine learning is to predict the mapping $f : \mathbf{q} \mapsto (\gamma, \Delta
\mathbf{\Lambda}, \mathbf{Q}) $, with which the corrected Reynolds stress $\mathbf{R}^{*} $ can be recovered as
follows:
\begin{equation}
  \label{eq:tau-prime}
  \mathbf{R}^{*} 
  = 2 k^* \left( \frac{\mathbf{I}}{3} + \mathbf{V}^* \mathbf{\Lambda}^* \mathbf{V}^{*T} \right) 
  = 2 \gamma_k k \left( \frac{\mathbf{I}}{3} + \mathbf{Q}\mathbf{V} (\mathbf{\Lambda} + \Delta \mathbf{\Lambda})
    \mathbf{V}^T \mathbf{Q}^T \right) 
\end{equation}
While representing the perturbations on the magnitude and the shape is straightforward, the rotation can
be difficult to represent. A few schemes are introduced and examined below.

\section{Representing rigid-body rotation of eigenvectors in machine learning}
\label{sec:metric}
It has been established above that representing eigenvector perturbations as rigid-body rotations
ensures orthonormality of the obtained eigenvectors and realizability of the perturbed Reynolds stresses. An
apparently straightforward representation of such a rotation is to use the matrix
$\mathbf{Q}$. However, it has nine elements (i.e., direction cosines) but only three intrinsic degrees
of freedom due to the orthonormality constraint $\mathbf{Q}^T \mathbf{Q} = \mathbf{I}$. Existing
machine learning algorithms are not suitable for learning quantities with hard constraints, since they
were developed mostly for commercial applications (e.g., product preferences of customers), where
hard constraints are uncommon. Building constraints into the learning problem is not
trivial. Hence, it is more desirable to use a formulation with the same number of \emph{independent}
variables as the number of intrinsic degrees of freedom. Two candidate representations, Euler angles
and unit quaternion, are introduced and compared in the context of machine learning.

Before presenting the two representations, we shall first put forward two desirable properties of
the regression function $f: \mathbf{q} \mapsto \Delta \mathbf{R} $ to be learned from training
data. First, the function must be smooth for it to have good generalization performance, i.e., the
predictive performance of the trained function on data unseen during the training. The word
``smooth'' shall be interpreted in liberal sense (i.e., the output $\Delta \mathbf{R}$ varies mildly in
the feature $\mathbf{q}$ space) rather than in a mathematical sense (i.e., the existence of arbitrarily
high order derivatives). The reason for such a requirement is that non-smooth functions are more
susceptible to overfitting caused by the noises in the training data, which would in turn diminish
the predicative capabilities of the trained functions~\cite{domingos2012few}. Second, the regression
function itself must be frame-independent. This is a requirement unique for constitutive modeling in
computational mechanics (including turbulence modeling)~\cite[see
e.g.,][]{pope1975general,popebook}, which is equally applicable to theory-based modeling and
data-driven modeling.

For flows considered in this work (incompressible flows free from any geometric discontinuities),
the mean flow features $\mathbf{q}$ are smooth spatially. Moreover, all chosen features are
frame-independent. Consequently, the two requirements outlined above on $f$ translates to those on
the output $\Delta \mathbf{R}$, i.e., it should be spatially smooth and frame-independent. Note that
it is possible that both the input and the output (e.g., $\mathbf{q}$ and $\Delta \mathbf{R}$) are
frame-dependent but the mapping $f$ is frame-independent, which is common in many analytical
constitutive relations. We use the linear eddy viscosity model $\textrm{dev}(\mathbf{R}) = 2\nu_t \mathbf{S}$ to
illustrate this subtle point, where $\nu_t$ is a scalar eddy viscosity field. Here, both
$\mathbf{R}$ and $\mathbf{S}$ are frame-dependent, but the mapping itself is
frame-independent. However, if this function is to be learned from data, the training data need to
be duplicated in a large number of rotated frames so that the machine learning algorithm can
discover the uniqueness of that function in all these frames, which dramatically increases computational
costs. Therefore, it is preferable to use the invariants of the tensors of concern ($\mathbf{S}$ and $\mathbf{R}$) 
as inputs and outputs instead~\cite{ling2016JCP}. This is a unique
complication in machine-learning-based modeling that is not present in traditional, theory-based
modeling.

\subsection{Euler angles}
The Euler angles used in this work follows the $z$--$x'$--$z''$ convention in rigid body
dynamics~\cite{goldstein1980euler}. That is, if a local coordinate system $x$--$y$--$z$ spanned by
the three eigenvectors was initially aligned with the global coordinate system ($X$--$Y$--$Z$), the
current configuration could be obtained by the following three consecutive intrinsic rotations about
the axes of the local coordinate system: (1) a rotation about the $z$ axis by angle $\phi_1$, (2) a
rotation about the $x$ axis by $\phi_2$, and (3) another rotation about its $z$ axis by $\phi_3$.
In general the local coordinate axes change orientations after each rotation.  Such a convention
provides a set of Euler angles $(\phi_1, \phi_2, \phi_3)$ to describe the current orientation of a rigid body (or
eigenvectors $\mathbf{V}$ in this case) within a global coordinate system. We refer to this
description as ``absolute Euler angles''.  With this description, the discrepancy between two sets of
eigenvectors, $\mathbf{V}$ and $\mathbf{V}^*$, can be described by the \emph{discrepancies
  $\Delta \phi_\alpha$ in their absolute Euler angles}, $\phi_\alpha$ and $\phi_\alpha^*$,
respectively, with $\Delta \phi_\alpha = \phi_\alpha^* - \phi_\alpha$ and $\alpha = 1, 2, 3$. In
the machine-learning-assisted turbulence modeling framework of Wang et al.~\cite{wang2017physics}, regression
functions for $\Delta \phi_\alpha$ were trained by using high-fidelity data.  However, it can be
seen that this representation relies on a global coordinate system, which makes it frame-dependent.
In particular, it is inadequate for more complex scenarios and can lead to deteriorated predictive
performances. The importance of frame-independence in machine-learning-assisted physical modeling
has been discussed in~\cite{ling2016JCP}.  A possible remedy of the frame dependence in the formulation
above is to directly describe the rotation from $\mathbf{V}$ to $\mathbf{V}^*$ in Euler angles
$\phi_\alpha^{o\to*}$, where $o$ and $*$ in the superscript indicate original and corrected,
respectively. However, as will be shown by the \textit{a priori} test in
Section~\ref{sec:a-priori}, the direct-rotation-based Euler angles $\phi_\alpha^{o\to*}$ are non-smooth
spatially and thus undesirable in light of the smoothness requirement explained above.

In summary, the two variants of Euler-angle-based representations of the perturbation from
$\mathbf{V}$ to $\mathbf{V}^*$, i.e., discrepancy-based representation $\Delta \phi_\alpha$ and
direct-rotation-based representation $\phi_\alpha^{o\to*}$, are plagued by their own weaknesses,
namely the frame-independence and non-smoothness, respectively. These difficulties prompted us to
explore the unit quaternion as an alternative representation.

\subsection{Unit quaternion}

Given two sets of orthonormal eigenvectors $\mathbf{V}$ and $\mathbf{V}^*$ sharing the same origin
$O$, the Euler's rotation theorem states there exists a unique axis of unit vector $\mathbf{n} \equiv
[n_1, n_2, n_3]$ and an angle $\theta$ such that $\mathbf{V}^*$ can be obtained via rotating
$\mathbf{V}$ by $\theta$ about an axis $\mathbf{n}$ that runs through the origin $O$. The rotation
can thus be represented compactly with a unit quaternion:
\begin{equation}
  \label{eq:quaternion}
  \mathbf{h} = \left[\cos{\frac{\theta}{2}}, \; n_{1}\sin{\frac{\theta}{2}}, \; n_{2}\sin{\frac{\theta}{2}}, \;  n_{3}\sin{\frac{\theta}{2}}\right]^T
\end{equation}
which clearly verifies $\|\mathbf{h}\| = 1$, with $\| \cdot \|$ indicating Euclidean-norm.  The axis
$\mathbf{n}$ and the angle $\theta$ are both determined by $\mathbf{V}$ and $\mathbf{V}^*$ and do
not depend on a global coordinate system. Therefore, the unit-quaternion-based representation of the
rotation is frame-independent. 

To conclude this section, we point out that any rotation that transforms $\mathbf{V}$ to $\mathbf{V}^*$ can be uniquely represented by any of the following:
\begin{enumerate}[(i)]
\item a rotation matrix $\mathbf{Q}$;
\item a unique set of Euler angles
$(\Delta \phi_1, \Delta \phi_2, \Delta \phi_3)$ or $(\phi_1^{o\to*}, \phi_2^{o\to*},\phi_3^{o\to*})$ based on discrepancy or direct rotation, respectively; and
\item a unit quaternion~$\mathbf{h}$.
\end{enumerate}
A rare exception is the scenarios involving gimbal lock for the Euler
angles~\cite{kuipers1999quaternions}.

\section{Numerical results}
\label{sec:results}
Numerical examples are used to evaluate the performances of three representations of perturbations
to Reynolds stress eigenvectors, i.e., (1) direct-rotation-based Euler angles, (2) discrepancy-based
Euler angles, and (3) unit quaternion. In the $\textit{a priori}$ tests, we compute the ``true''
perturbations that are needed to correct the RANS modeled Reynolds stress to the DNS results by
using these three representations. The smoothness of such true perturbation fields is indicative of
the difficulties when using data to learn the regression functions for the perturbations. The tests
show that latter two representations (discrepancy-based Euler angles and unit quaternion) satisfy
the smoothness requirement.  Therefore, in $\textit{a posteriori}$ tests we focus on these two
representations.  Predictive performances of the machine-learning models with the later two
representations are assessed by investigating several training-prediction scenarios.  The results suggest that
the unit quaternion representation leads to better results due to its frame-independence. This
advantage is particularly clear when the training and test flows involve different coordinate
systems or geometries.

\subsection{Simulation setup}
\begin{figure}[htbp]
\centering
\includegraphics[width=0.7\textwidth]{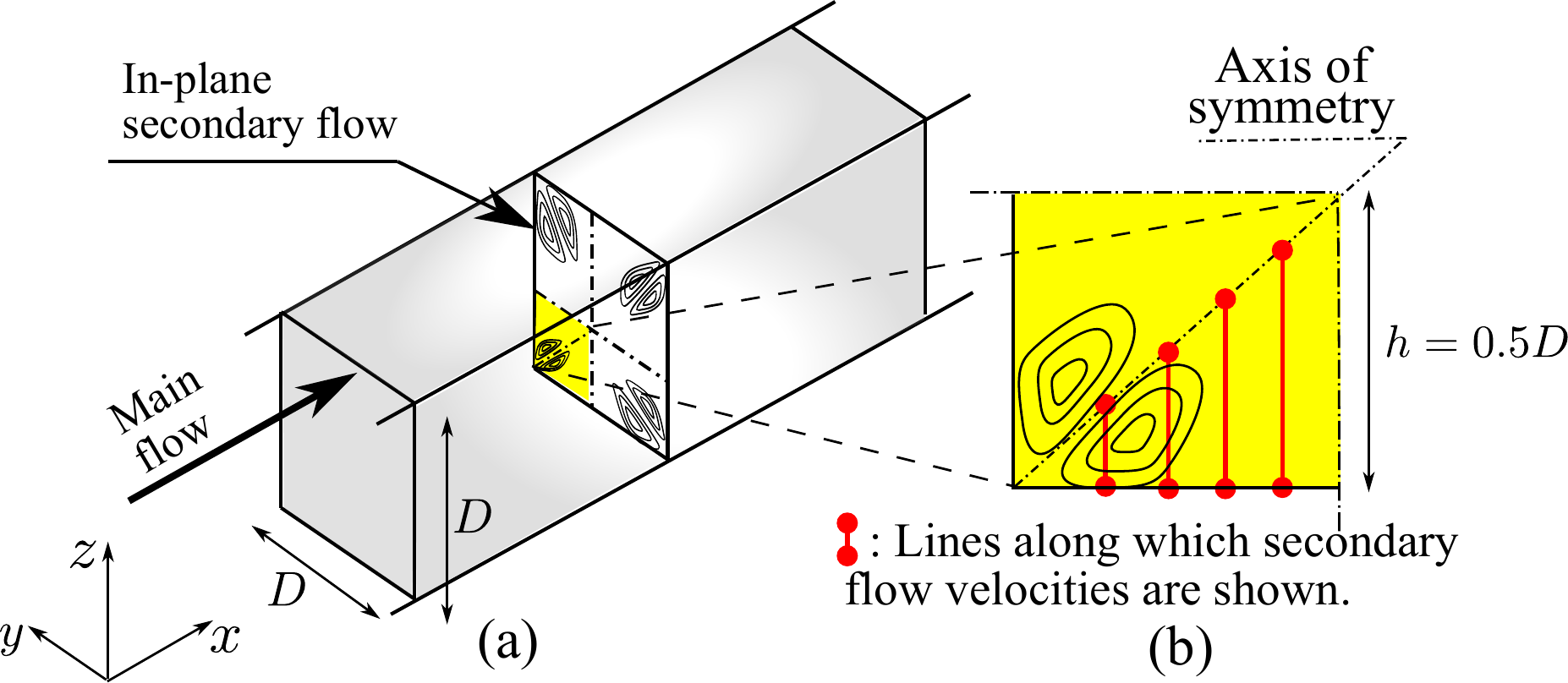}
\caption{Computational domain for the flow in a square duct. The $x$ coordinate represents the streamwise
  direction. Secondary flows induced by Reynolds stress imbalance exist in the $y$--$z$
  plane as shown in panel (a). Panel (b) shows that the computational domain covers a quarter of the cross-section of the
  physical domain.}
\label{fig:domain_duct}
\end{figure}

In this work, two test cases are employed to compare the performances of three representations of eigenvectors
perturbations on Reynolds stress eigenvectors. The first test case is fully developed turbulent flow in a square duct. It is well known that RANS models have difficulty in predicting the secondary flow induced by Reynolds stress imbalances~\cite{huser1993direct}.  A schematic is presented in Fig.~\ref{fig:domain_duct} to show the physical domain and the computational domain. A two-dimensional simulation is performed, since the flow is fully developed along the stream-wise direction. In addition, the computational domain only covers a quarter of the cross-section as shown in Fig.~\ref{fig:domain_duct}b due to the symmetry of the flow along $y$ and $z$ directions. All lengths are normalized by the height of the computational domain $h=0.5D$, where $D$ is the height of the duct. The Reynolds number $Re$ is based on the height of the computational domain $h$ and bulk velocity $U_b$. 


The RANS simulations are performed in an open-source CFD platform OpenFOAM, using a
built-in steady-state incompressible flow solver \texttt{simpleFoam}~\citep{weller1998tensorial}, in
which the SIMPLE algorithm~\cite{patankar1980numerical} is used. Launder-Gibson Reynolds stress
transport model~\cite{gibson1978ground} is used for the RANS simulations of both the training
flow and the test flow. In the RANS simulations, the $y^+$ of the first cell center is kept less
than 1 and thus no wall model is applied.  DNS data of the training flows are obtained from Pinelli
et al.~\cite{pinelli2010reynolds}.

\begin{figure}[htbp]
\centering
\includegraphics[width=0.8\textwidth]{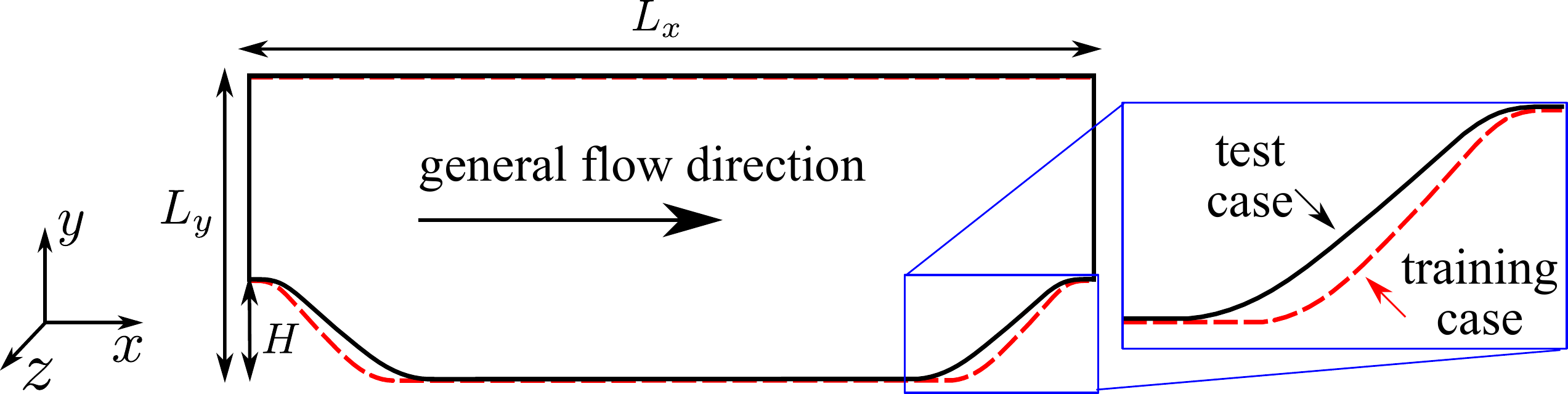}
\caption{Computational domain for the flow over periodic hills. The solid line indicates the configuration of the test flow. A zoom-in view shows the comparison between the hill profiles of the training flow (dashed line) and the test flow (solid line). The hill width of the training flow is 0.8 of the hill width of the test flow. The $x$, $y$ and $z$ coordinates
  are aligned with streamwise, wall-normal, and spanwise, respectively.}
\label{fig:domain_pehill}
\end{figure}

Another training--prediction scenario consists of the flows over periodic hills as shown in Fig.~\ref{fig:domain_pehill}. The test flow is the flow over periodic hills at $Re=5600$.
The geometry of the computational domain of the test flow is shown in Fig.~\ref{fig:domain_pehill}. Compared to the test flow, the training flow is at the same Reynolds number but with a steeper hill profile as shown in Fig.~\ref{fig:domain_pehill} indicated by the dashed line. The Reynolds number 
$Re$ is based on the crest height $H$ and the bulk velocity $U_b$ at the crest. Periodic boundary conditions are applied in the streamwise ($x$) direction, 
and non-slip boundary conditions are applied at the walls. Both the DNS simulations of the training flow and the test flow are performed by using Incompact3d~\cite{laizet2011incompact3d}. In Incompact3d, the incompressible Navier--Stokes equations are solved on a Cartesian mesh using sixth-order finite-difference compact schemes for the spatial discretisation and a third-order Adams--Bashforth scheme for the time advancement. More details about the numerical methods used in Incompact3d can be found in~\cite{laizet2009high}. A validation of our DNS results of velocity field and Reynolds stress components of the test flow show a good agreement with the results in literature~\cite{breuer2008}. All the baseline RANS simulations used Launder-Sharma $k$-$\varepsilon$ model~\cite{launder1974application}. The $y^+$ of the first cell center is kept below 1, and thus no wall model is applied.

All the training-prediction scenarios in the \textit{a posteriori} tests are summarized in Table.~\ref{tab:scenarios}. The scenario 1 is investigated to show the ideal scenario where the prediction performances based on discrepancy-based Euler angles and unit quaternion are comparable to each other. The scenarios 2 and 3 are chosen to demonstrate the potential insufficiency of using discrepancy-based Euler angles in complex flow problems. 
\begin{table}[htbp] 	
  \centering
  \caption{The training-prediction scenarios in $\textit{a posteriori}$ test.
}
\label{tab:scenarios}
\begin{tabular}{P{2.0cm} | P{6.0cm}  P{6.0cm} }	
  \hline
 Cases  & Training set & Test set \\ 
  \hline
  1  & Flow in a square duct at $Re=2900$~\cite{pinelli2010reynolds} & Flow in a square duct at $Re=3500$~\cite{pinelli2010reynolds} \\  
  \hline	
  2  & Flow in a square duct at $Re=2900$ (coordinate system rotated anti-clockwise by $60 \degree$) & Flow in a square duct at $Re=3500$~\cite{pinelli2010reynolds} \\  
  \hline	
  3  & Flow over periodic hills at $Re=5600$ (steeper hill profile) & Flow over periodic hills at $Re=5600$~\cite{breuer2009flow} \\  
  \hline					 								
\end{tabular}
\flushleft
\end{table}

\subsection{A priori results}
\label{sec:a-priori}
\begin{figure}[!htbp]
  \centering
  {\includegraphics[width=0.35\textwidth]{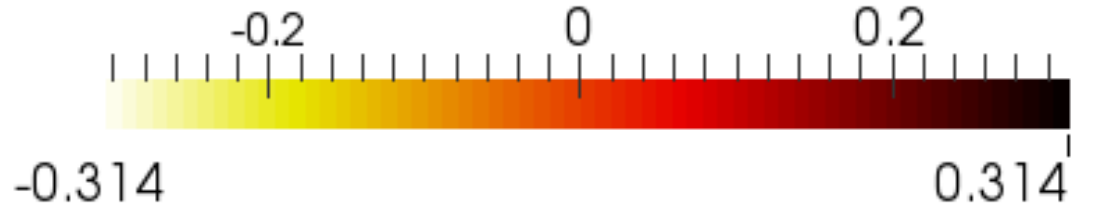}}\\
  \subfloat[$\phi_1^{o\to*}$]{\includegraphics[width=0.25\textwidth]{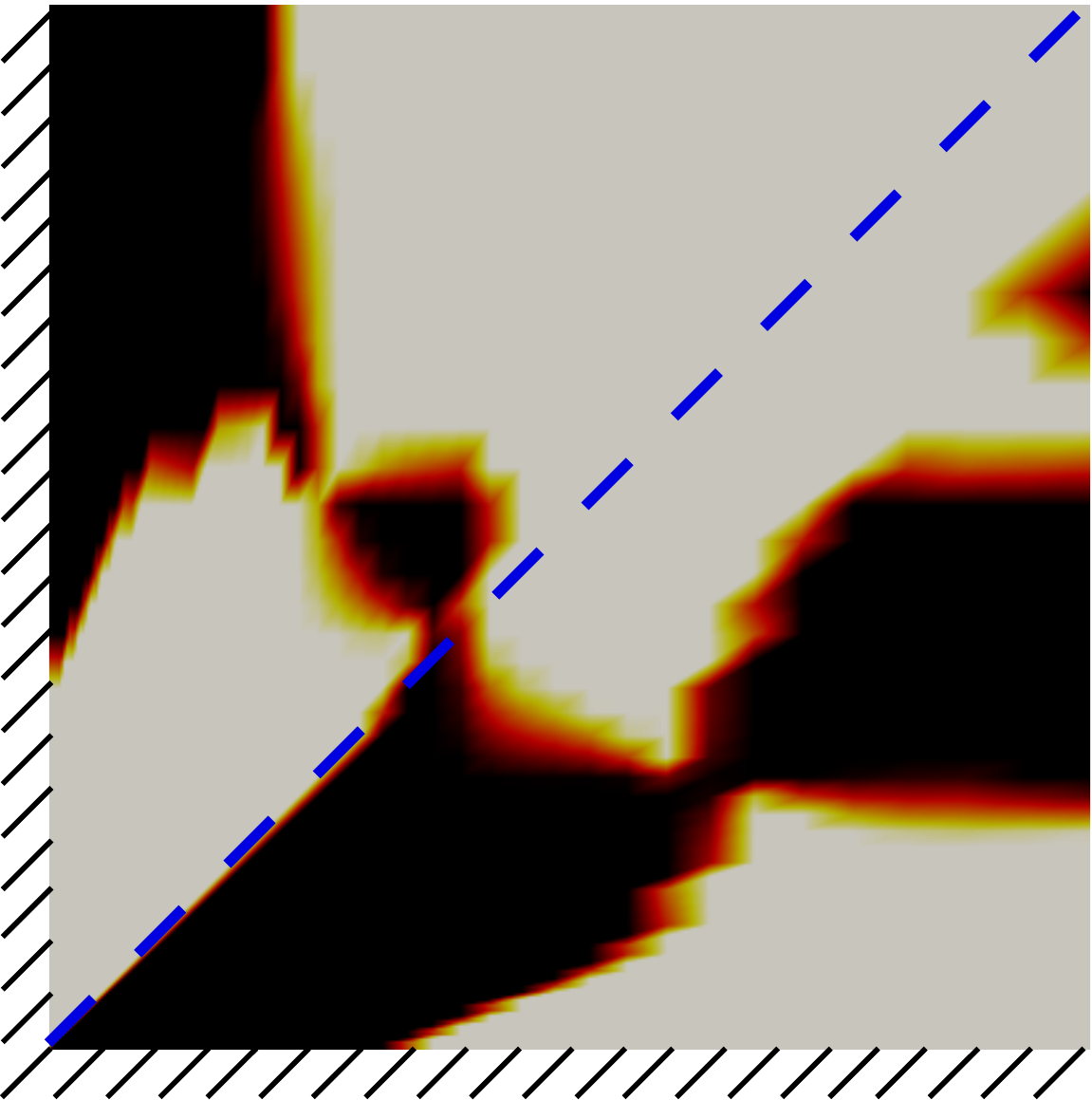}}\
  \subfloat[$\phi_2^{o\to*}$]{\includegraphics[width=0.25\textwidth]{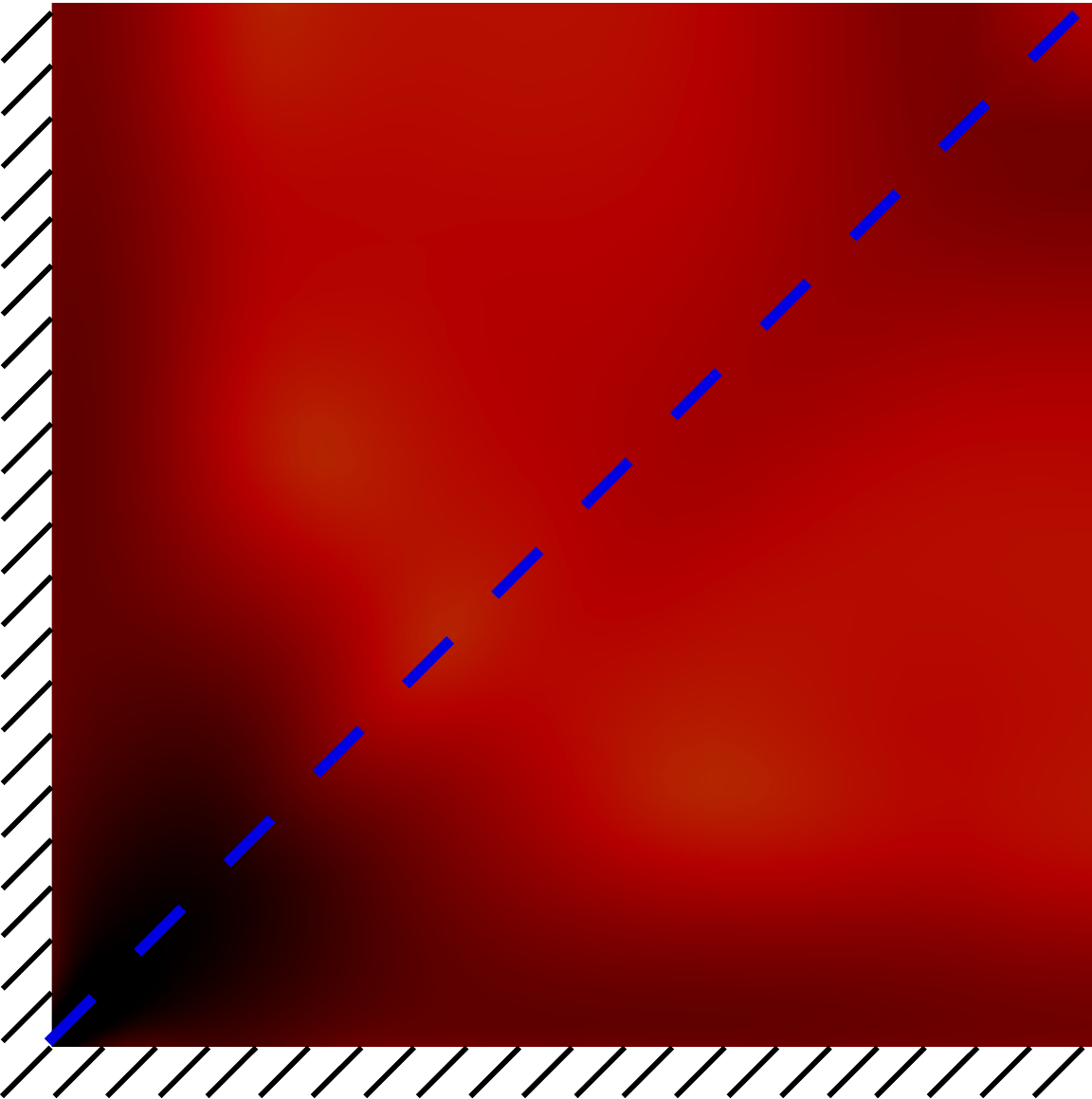}}\
  \subfloat[$\phi_3^{o\to*}$]{\includegraphics[width=0.25\textwidth]{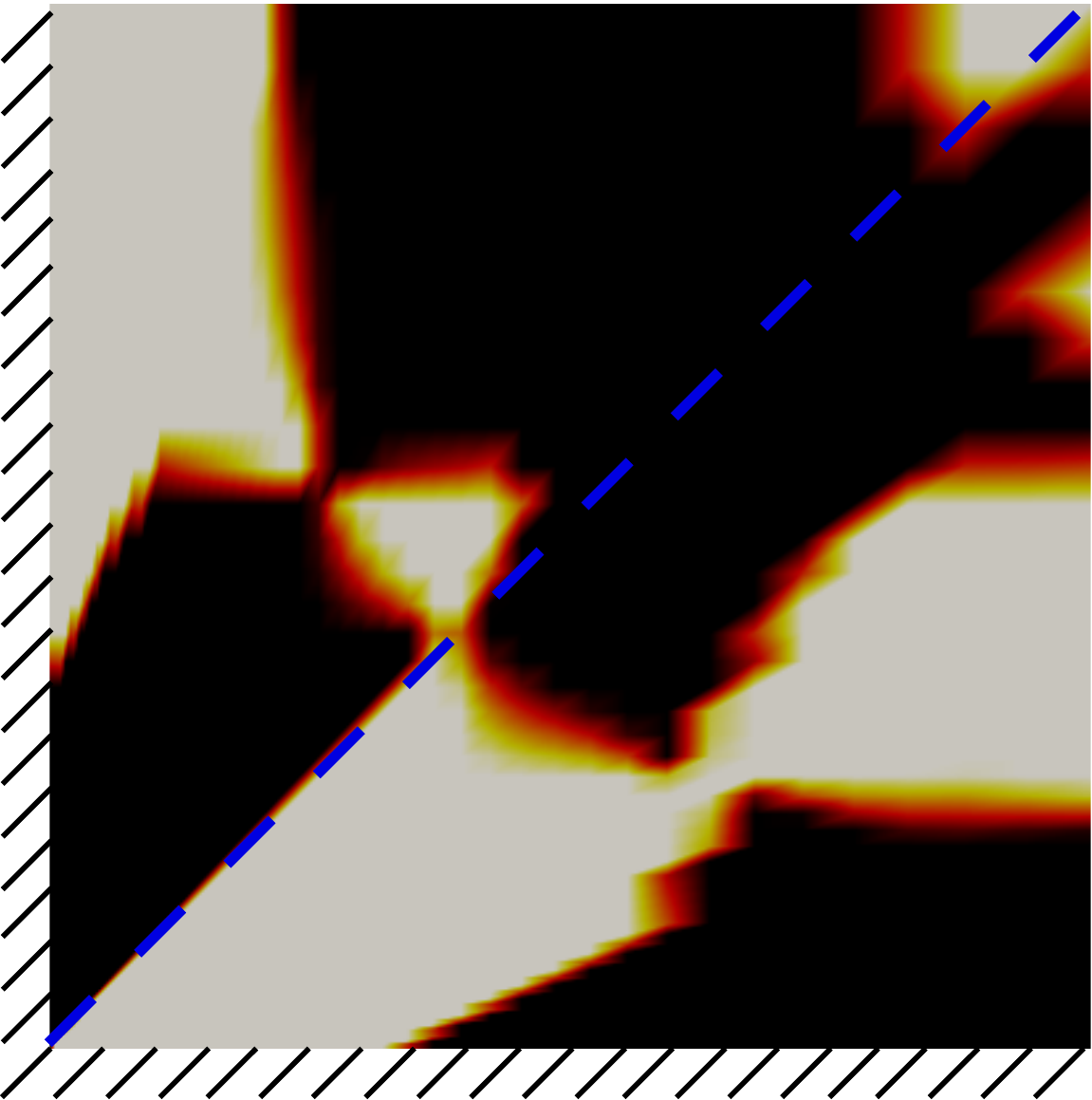}}\\
  \subfloat[$\Delta \phi_1$]{\includegraphics[width=0.25\textwidth]{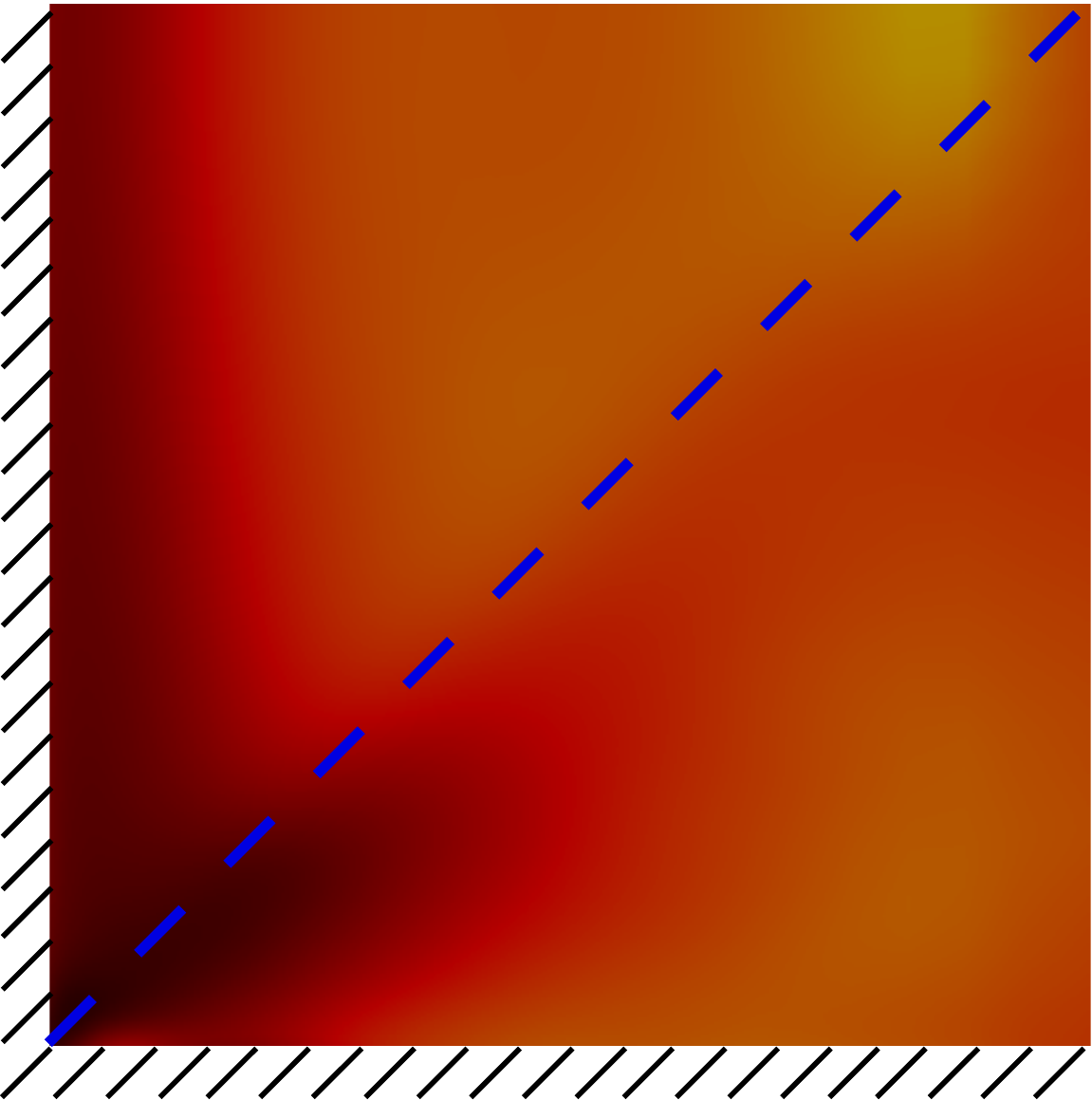}}\
  \subfloat[$\Delta \phi_2$]{\includegraphics[width=0.25\textwidth]{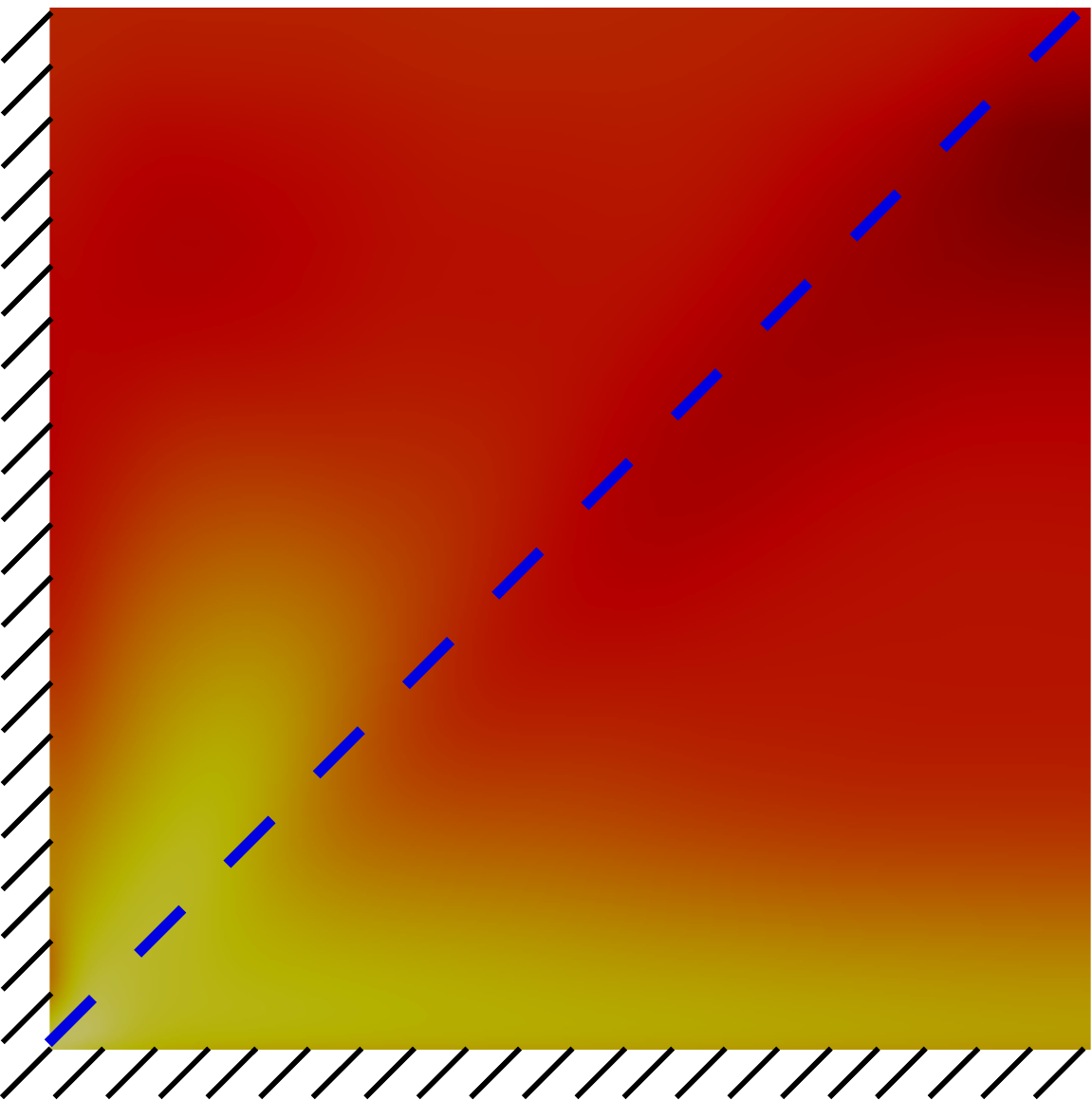}}\
  \subfloat[$\Delta \phi_3$]{\includegraphics[width=0.25\textwidth]{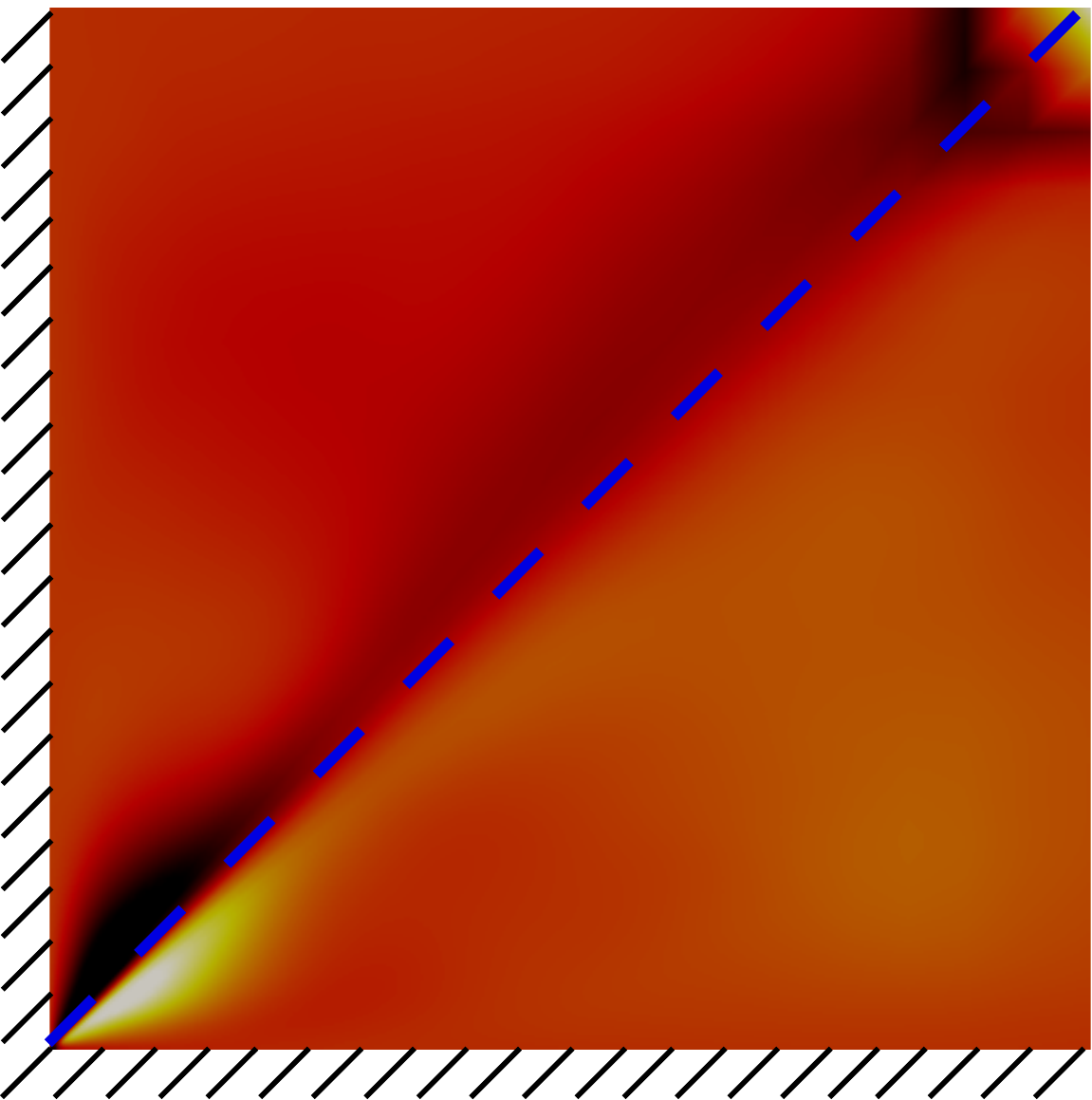}}\
  \caption{Desired perturbations between the eigenvectors of the RANS modeled Reynolds stress and that of the DNS data, represented by direct-rotation-based Euler angles $\phi_{\alpha}^{o\to*}$ (top row) and discrepancy-based Euler angles $\Delta \phi_{\alpha}$ (bottom row).  The dashed line denotes the diagonal of the square duct.}
\label{fig:euler-priori}
\end{figure}

In this $\textit{a priori}$ test, we first demonstrate that the discrepancy-based Euler angles are spatially smooth, while the direct-rotation-based Euler angles are not. It can be seen in Figs.~\ref{fig:euler-priori}a--\ref{fig:euler-priori}c that the direct-rotation-based Euler angles lack the smoothness, particularly for the angles $\phi_1^{o\to*}$ and $\phi_3^{o\to*}$. Such lack of smoothness of Euler angles $\phi_1^{o\to*}$ and $\phi_3^{o\to*}$ can be explained by the rotation matrix $\mathbf{Q}$. Specifically, the angles $\phi_1^{o\to*}$ and $\phi_3^{o\to*}$ are determined by the ratio between off-diagonal terms of the rotation matrix $\mathbf{Q}$. In the scenario that the two eigenvectors systems are close to each other, it is expected that the rotation matrix $\mathbf{Q}$ would be diagonal dominant, and the off-diagonal terms should be small. However, the ratio between the off-diagonal terms are not necessarily small, and such ratio would be more sensitive to the change of off-diagonal Reynolds stress components. It explains the lack of smoothness for values of $\phi_1^{o\to*}$ and $\phi_3^{o\to*}$. On the other hand, it can be seen from Figs.~\ref{fig:euler-priori}d--\ref{fig:euler-priori}f that the discrepancy-based Euler angles are smoother than the original Euler angles. The main reason is that the discrepancy-based Euler angles are obtained based on the difference between two sets of direct-rotation-based Euler angles , i.e., one set from RANS modeled Reynolds stress and the other from DNS data, with respect to the same global reference frame. Thus, the effect of sensitivity issue due to the small off-diagonal components of Reynolds stress tends to be canceled out, and better spatial smoothness is achieved.

\begin{figure}[!htbp]
  \centering
  \subfloat[$h_1$]{\includegraphics[width=0.25\textwidth]{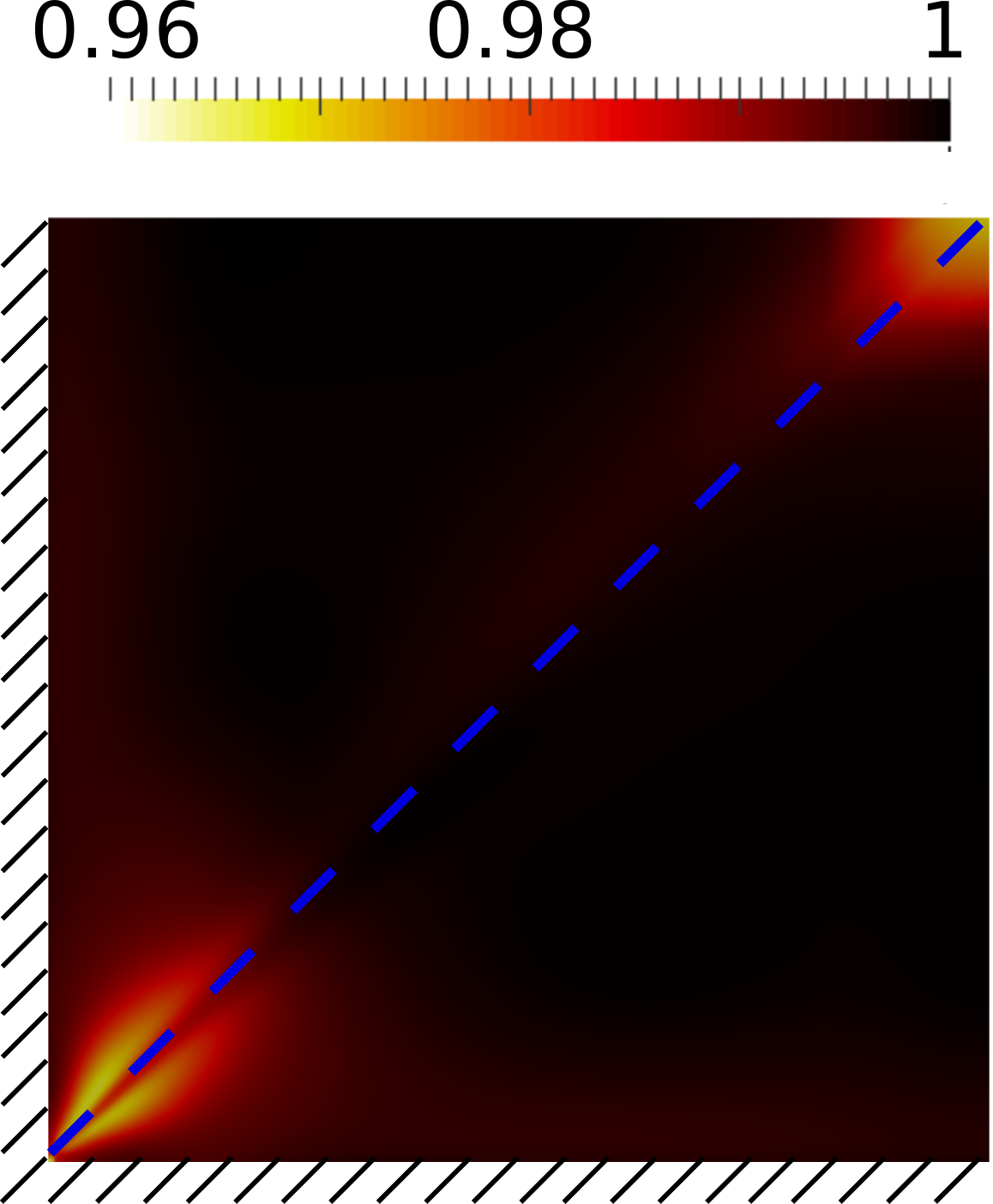}}\
  \subfloat[$h_2$]{\includegraphics[width=0.25\textwidth]{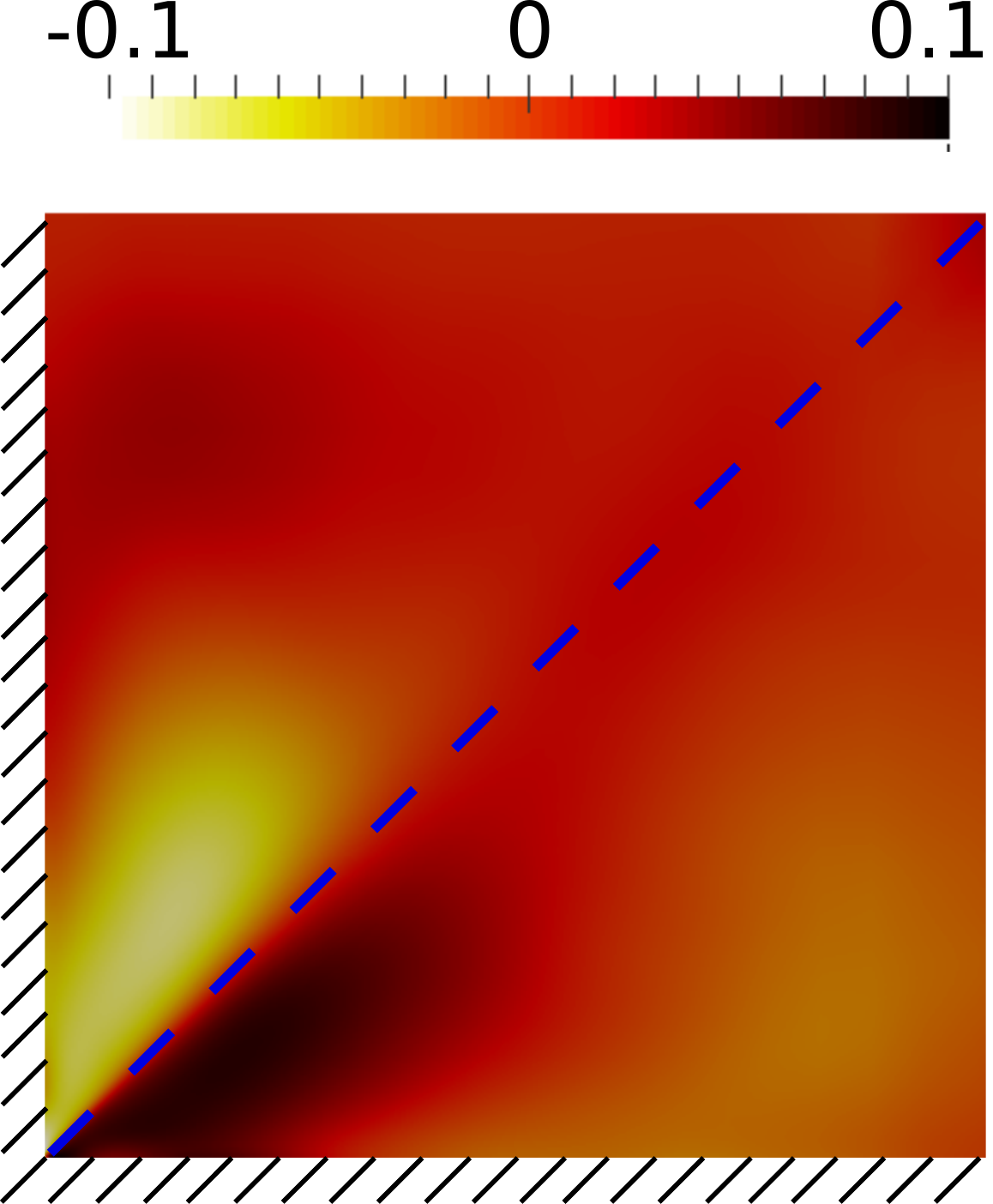}}\
  \subfloat[$h_3$]{\includegraphics[width=0.25\textwidth]{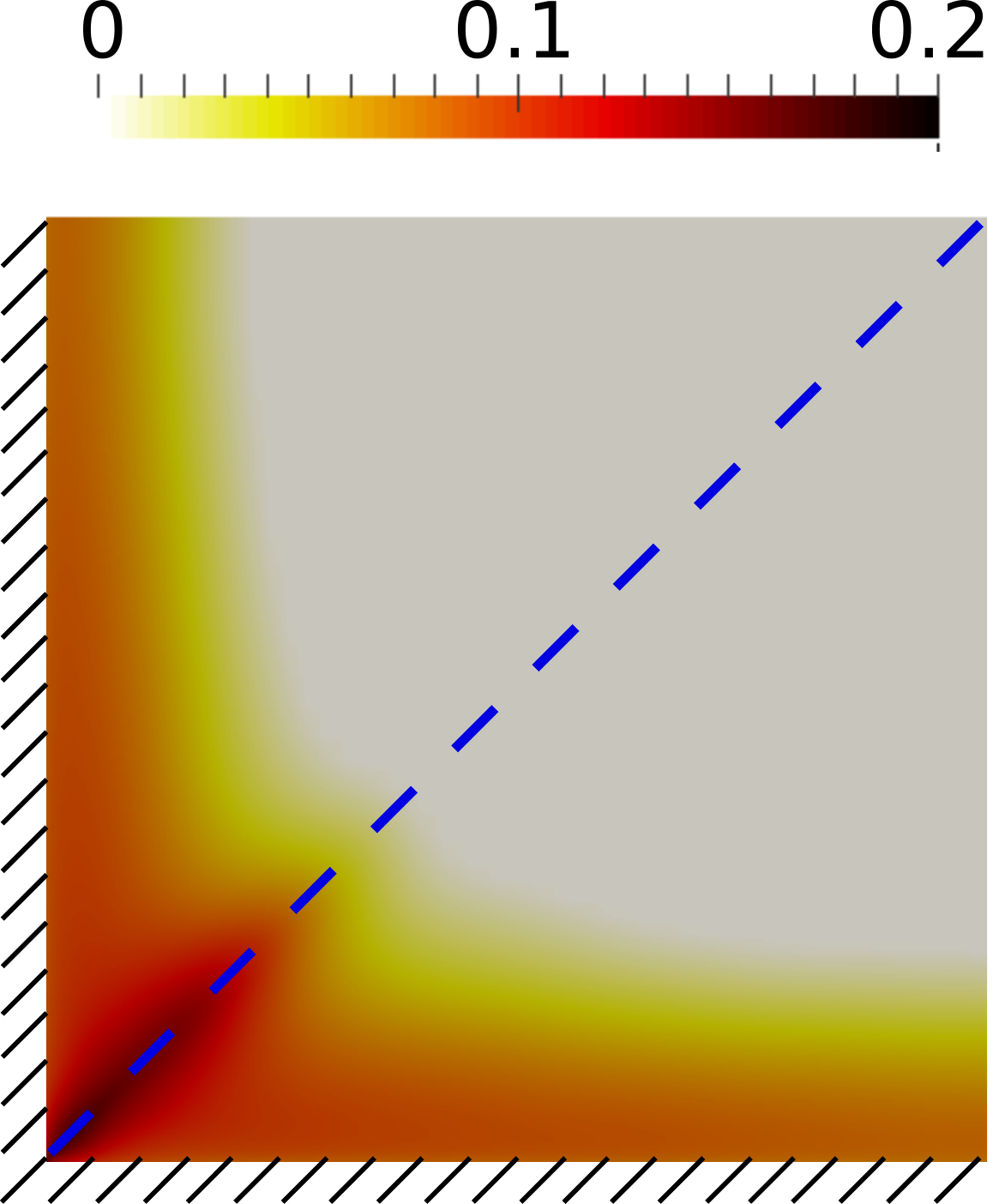}}\
  \caption{The desired perturbation between the eigenvectors of the RANS modeled Reynolds stress and that of the DNS data, represented by unit quaternion $\mathbf{h}$. The component $h_4$ is omitted here since it can be expressed in terms of $h_1$, $h_2$, and $h_3$. The dashed line denotes the diagonal of the square duct.}
\label{fig:quaternion-priori}
\end{figure}

The spatial smoothness can also be achieved in most regions as shown in Fig.~\ref{fig:quaternion-priori} by using unit quaternion to represent the eigenvectors perturbations. It should be noted that the spatial smoothness is not achieved in Fig.~\ref{fig:quaternion-priori}b near the diagonal of square duct as indicated by dashed line. However, there are mean flow features within the invariant set of $\{\mathbf{S},\mathbf{\Omega},\nabla p,\nabla k\}$ with the similar antisymmetric pattern due to the rotational invariance. Therefore, the smoothness of the functional form $f:\, \mathbf{q} \mapsto \Delta \mathbf{R}$ can still be guaranteed. In addition, all the four components of unit quaternion show either symmetric or anti-symmetric pattern along the diagonal of the square duct denoted by the dashed line, indicating that these four components are invariants under the rotation of reference frame. It demonstrates the main advantage of unit quaternion compared to the discrepancy-based Euler angles shown in the bottom row of Fig.~\ref{fig:euler-priori}. More detailed comparisons between these two representations are presented in the $\textit{a posteriori}$ tests below.

\subsection{A posteriori results}

\label{sec:a-posteriori}
\begin{figure}[!htbp]
  \centering
  {\includegraphics[width=0.3\textwidth]{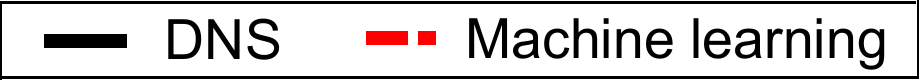}}\\
  \subfloat[$\Delta \phi_1$]{\includegraphics[width=0.32\textwidth]{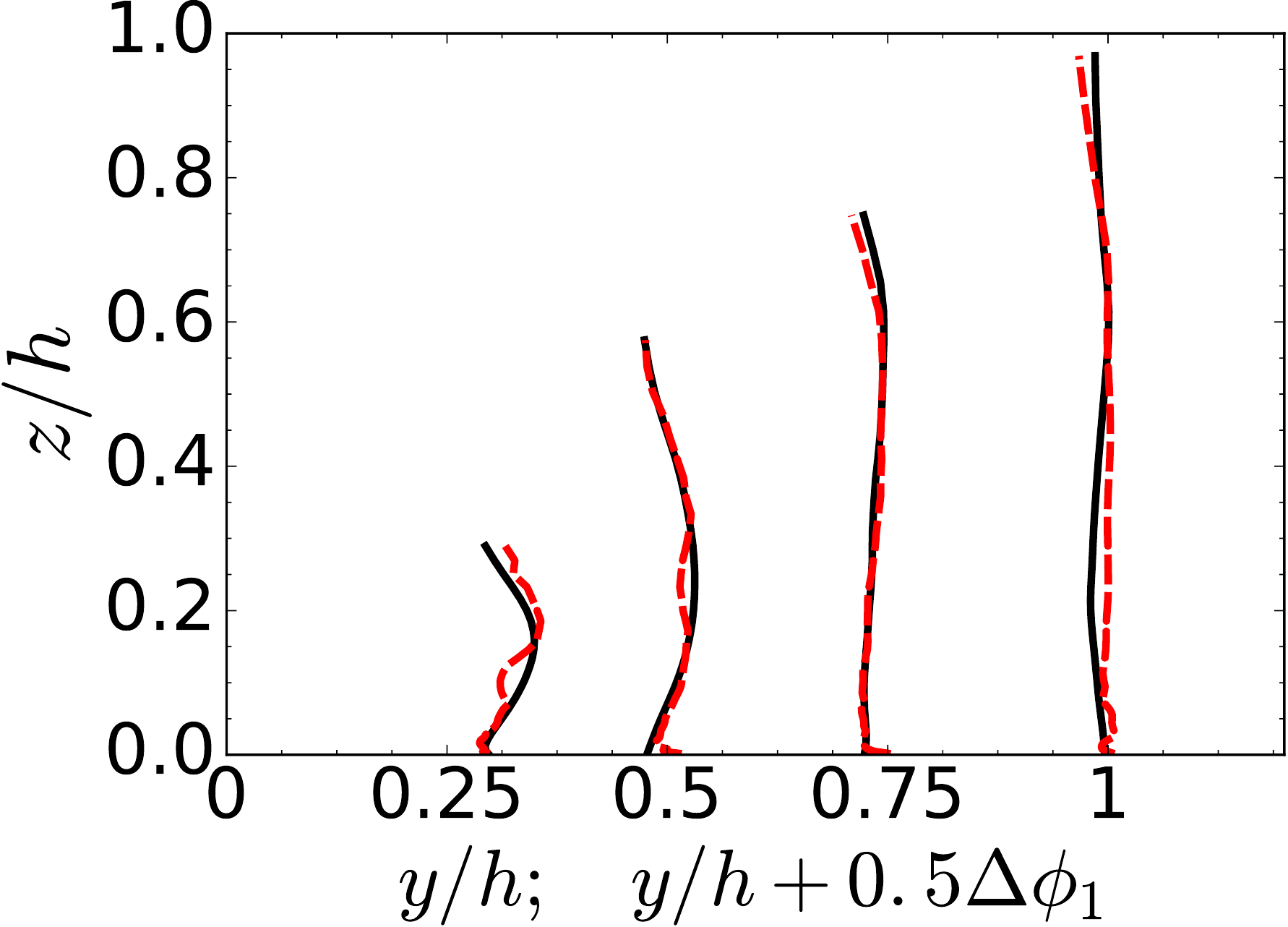}}\hspace{0.1em}
  \subfloat[$\Delta \phi_2$]{\includegraphics[width=0.32\textwidth]{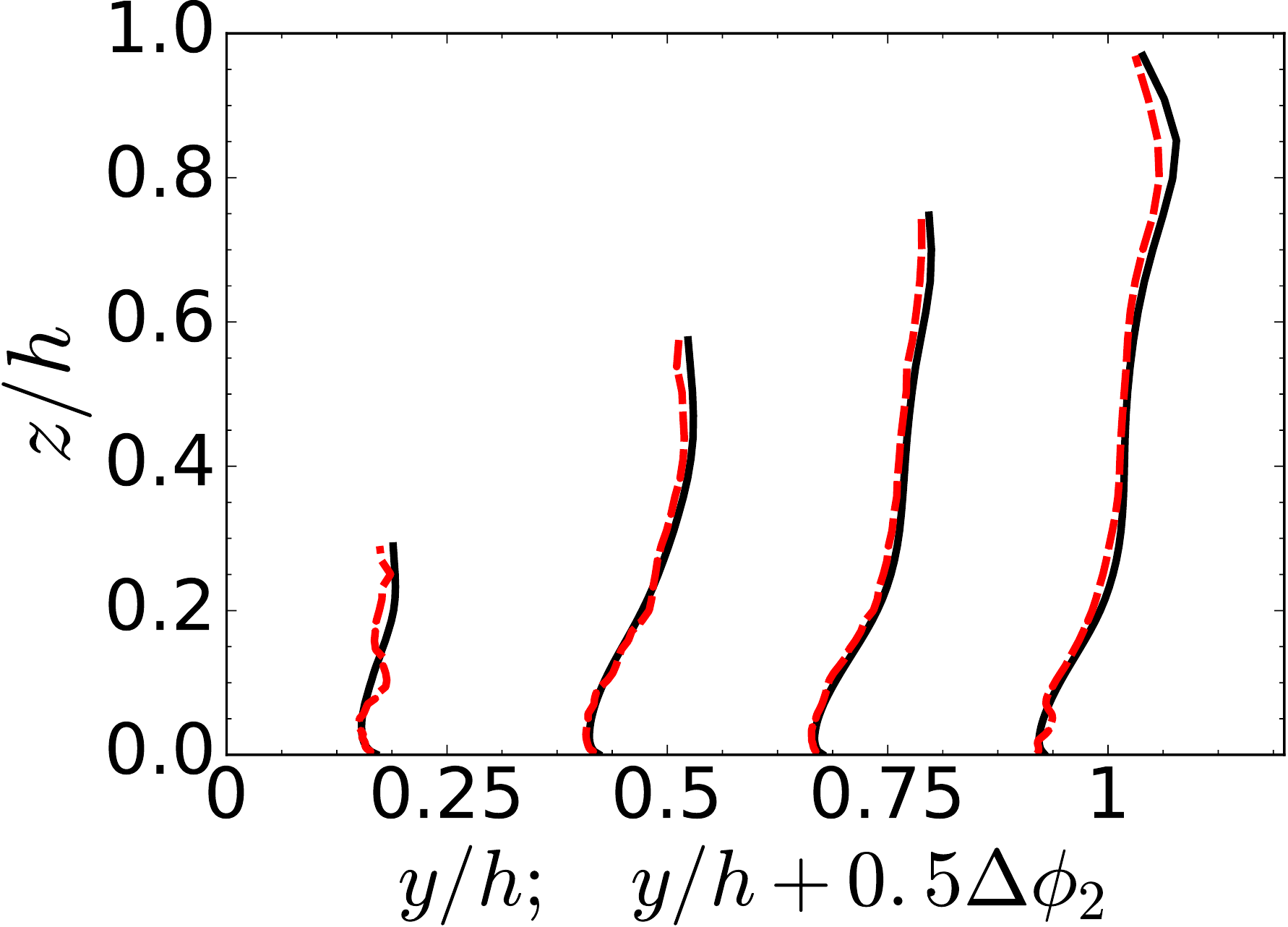}}\hspace{0.1em}
  \subfloat[$\Delta \phi_3$]{\includegraphics[width=0.32\textwidth]{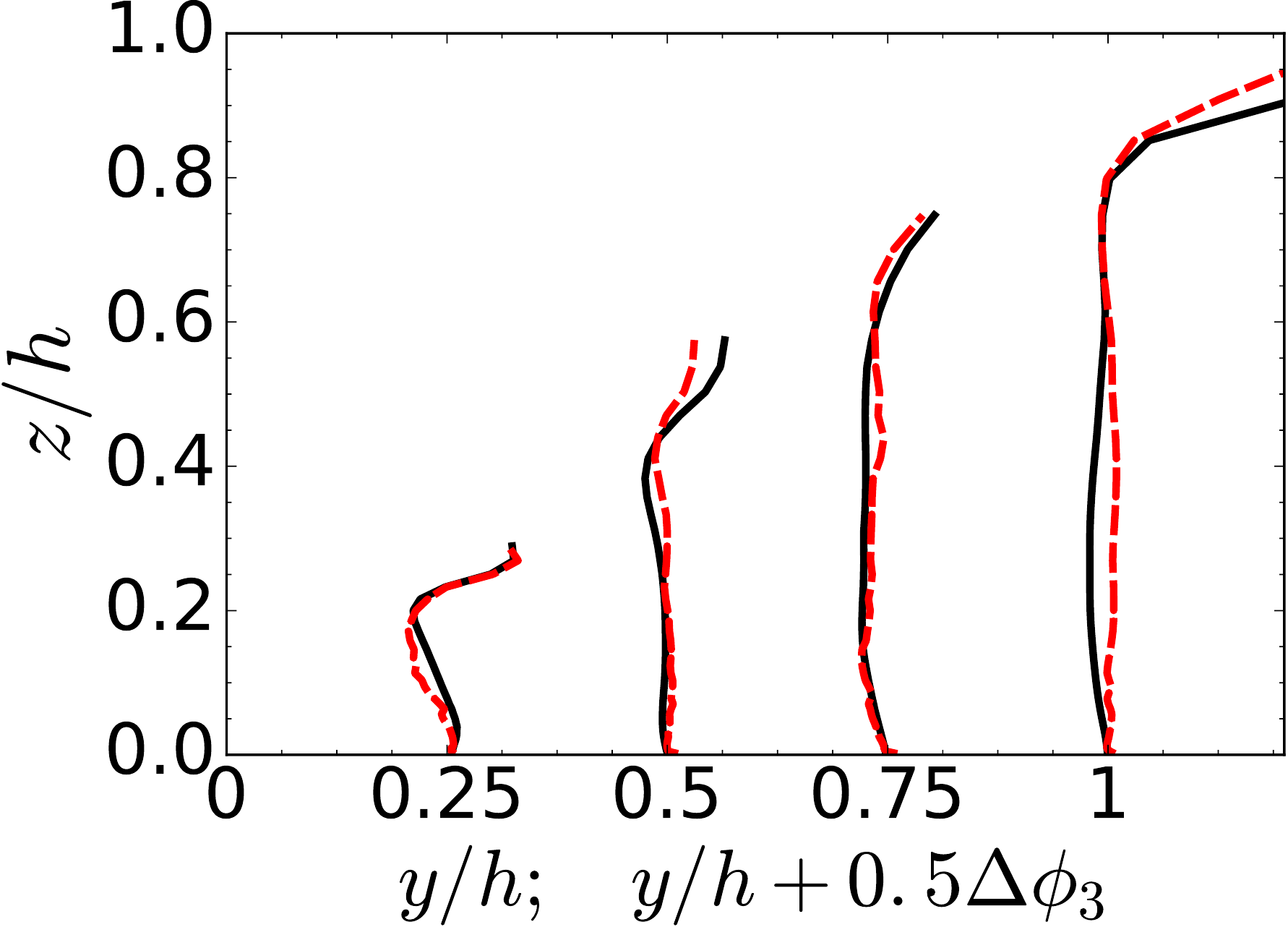}}
  \caption{The prediction of \textbf{discrepancy-based of Euler angles} for the flow in a square duct at $Re=3500$, showing three components: (a) $\Delta \phi_1$, (b) $\Delta \phi_2$ and (c) $\Delta \phi_3$. The training flow is the flow in a square duct at $Re=2900$.}
\label{fig:euler}
\end{figure}

In the posteriori tests, we investigate three training-prediction scenarios to demonstrate the merit of unit quaternion by comparing the machine learning performances of discrepancy-based Euler angles and the unit quaternion. In the first scenario, the flow in a square duct at $Re=3500$ is predicted by using the flow at a lower Reynolds number
$Re=2900$ as the training case. We first present the prediction of discrepancy-based Euler angles in Fig.~\ref{fig:euler}. It can be seen that the machine-learning-predicted discrepancy-based Euler angles demonstrate good agreements with the desired discrepancy-based Euler angles that perturb the eigenvectors of RANS modeled Reynolds stress to that of DNS data. The predictions of the desired perturbations of the eigenvalues and the turbulent kinetic energy (TKE) achieve the similar quality and thus are
omitted here, considering that this work focuses on the perturbations of the eigenvectors.

\begin{figure}[!htbp]
  \centering
  {\includegraphics[width=0.4\textwidth]{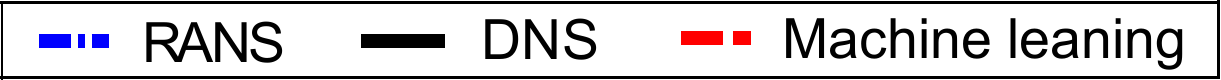}}\\
  \subfloat[$R_{xy}$]{\includegraphics[width=0.4\textwidth]{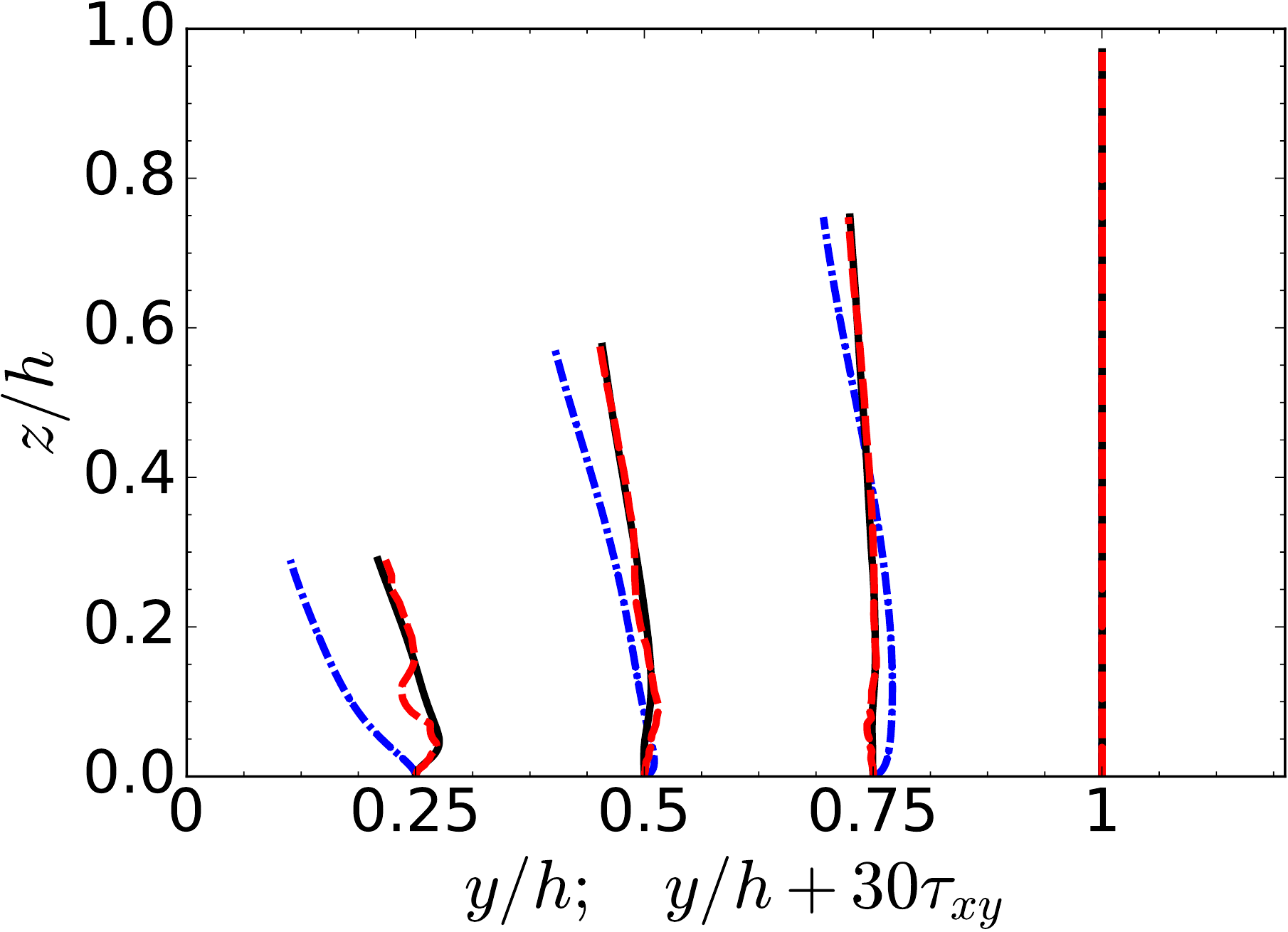}}\hspace{0.2em}
  \subfloat[$R_{xz}$]{\includegraphics[width=0.4\textwidth]{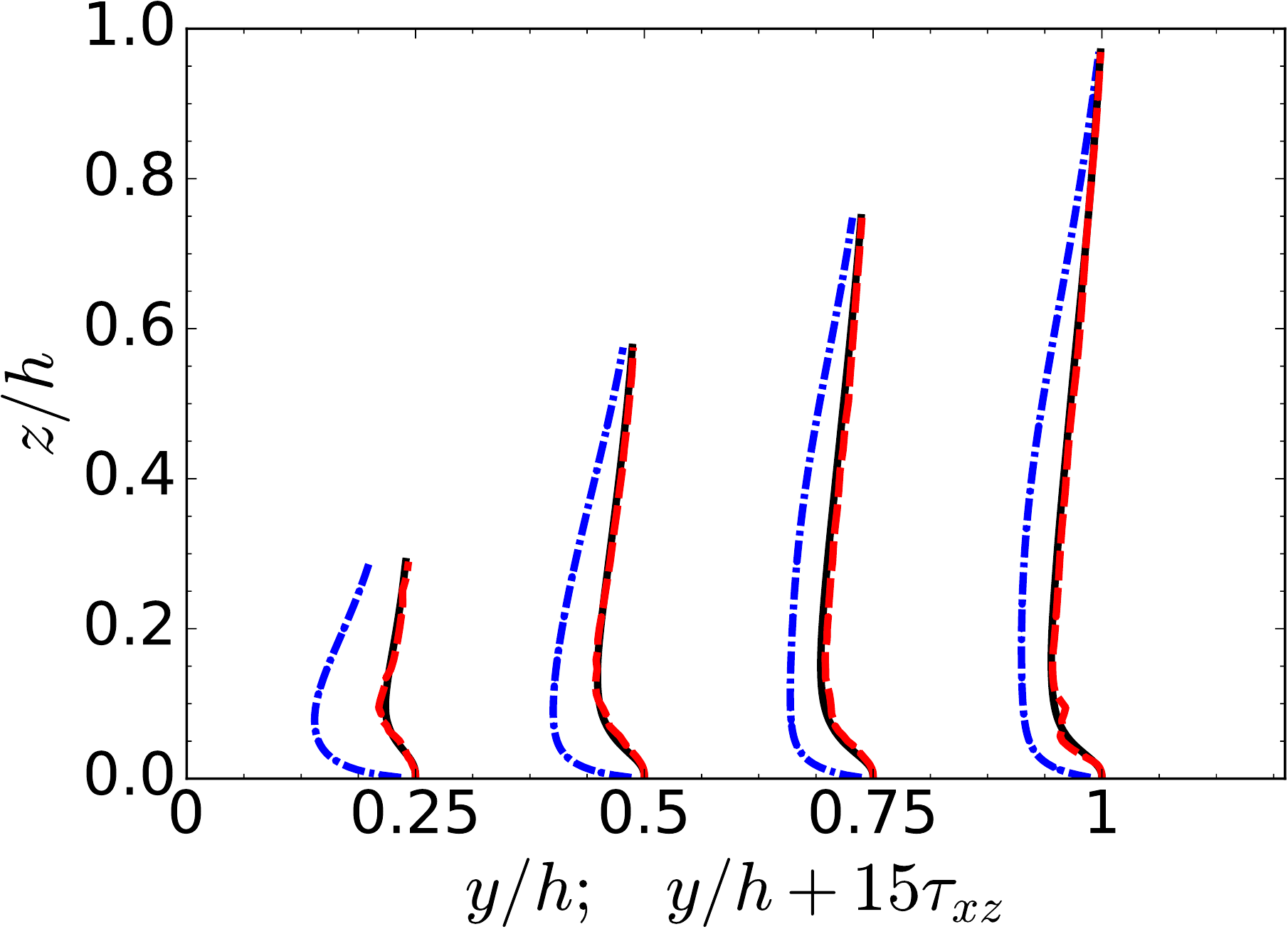}}
  \caption{The prediction of Reynolds stress components (a) $R_{xy}$ and (b) $R_{xz}$ based on \textbf{discrepancy-based Euler angles} for the flow in a square duct at $Re=3500$. The training flow is the flow in a square duct at $Re=2900$.}
\label{fig:euler-Tau}
\end{figure}

The reconstructed Reynolds stress components $R_{xy}$ and $R_{xz}$ based on the machine learning prediction of discrepancy-based Euler angles are shown in Fig.~\ref{fig:euler-Tau}. It should be noted that the machine-learning-predicted perturbations of the eigenvalues and the TKE are also employed in reconstructing the Reynolds stress components in all the following $\textit{a posteriori}$ tests. As shown in Fig.~\ref{fig:euler-Tau}a, the Reynolds stress component $R_{xy}$ is significantly overestimated by the baseline RANS simulation, especially at the near corner region. The baseline RANS modeled Reynolds stress component $R_{xz}$ is not satisfactory either, particularly near the corner. Compared to the 
baseline RANS results, both the reconstructed components $R_{xy}$ and $R_{xz}$ show much better 
agreements with the DNS data. This satisfactory prediction performance demonstrates the capability of using
discrepancy-based Euler angles in this specific training-prediction scenario, where the training case and the prediction case share the same geometry configuration and coordinate system.

\begin{figure}[!htbp]
  \centering
  {\includegraphics[width=0.3\textwidth]{component-legend}}\\
  \subfloat[$h_1$]{\includegraphics[width=0.32\textwidth]{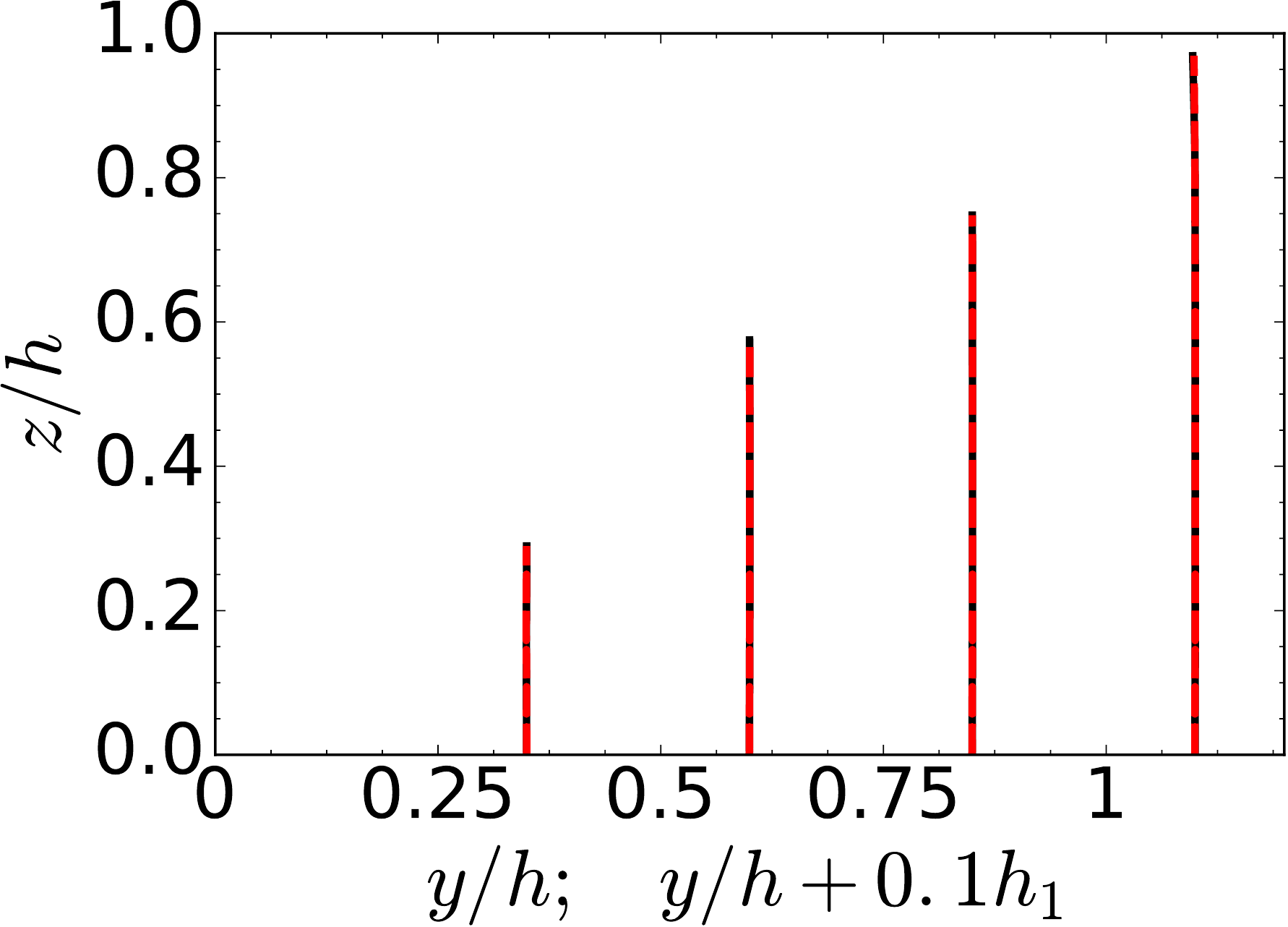}}\hspace{0.1em}
  \subfloat[$h_2$]{\includegraphics[width=0.32\textwidth]{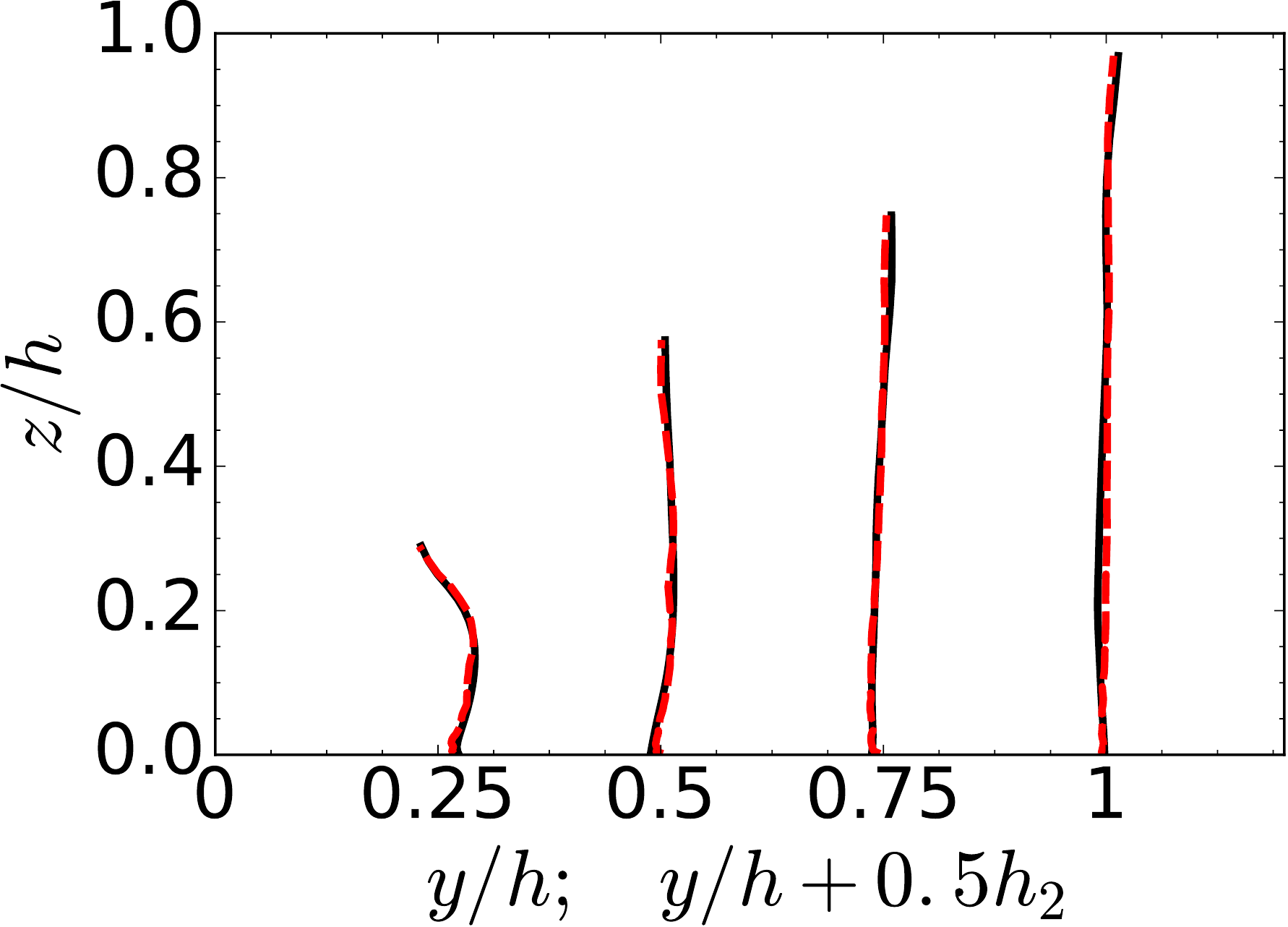}}\hspace{0.1em}
  \subfloat[$h_3$]{\includegraphics[width=0.32\textwidth]{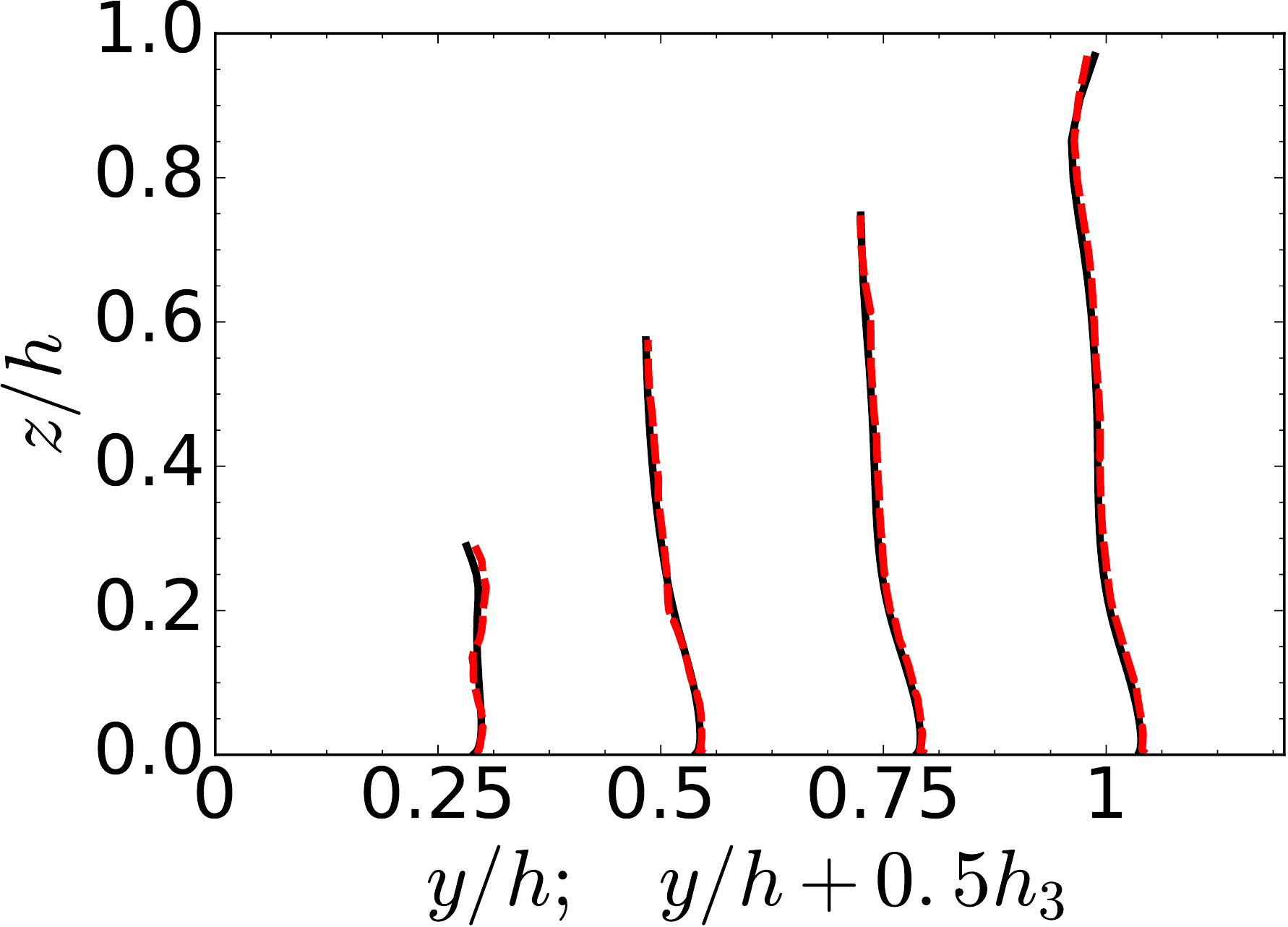}}
  \caption{The prediction of the components of unit quaternion for the flow in a square duct at $Re=3500$. The training flow is the flow in a square duct at $Re=2900$. The component $h_4$ is omitted here since unit quaternion only has three degrees of freedom.}
\label{fig:quaternion}
\end{figure}

In the first training-prediction scenario, we also tested the use of the unit quaternion representation. The results of unit quaternion are shown in Fig.~\ref{fig:quaternion}. It can be seen that the prediction performance of unit 
quaternion is similar to the prediction performance of discrepancy-based Euler angles as shown in Fig.~\ref{fig:euler}. It is because that the same geometry configuration and 
coordinate system are applied to both the training flow and the test flow. Thus, the frame-dependence of discrepancy-based Euler angles would not introduce additional errors into the machine learning prediction in this training-prediction scenario, explaining the similar prediction performance based on discrepancy-based Euler angles and unit quaternion.

\begin{figure}[!htbp]
  \centering
  {\includegraphics[width=0.4\textwidth]{Tau-legend}}\\
  \subfloat[$R_{xy}$]{\includegraphics[width=0.4\textwidth]{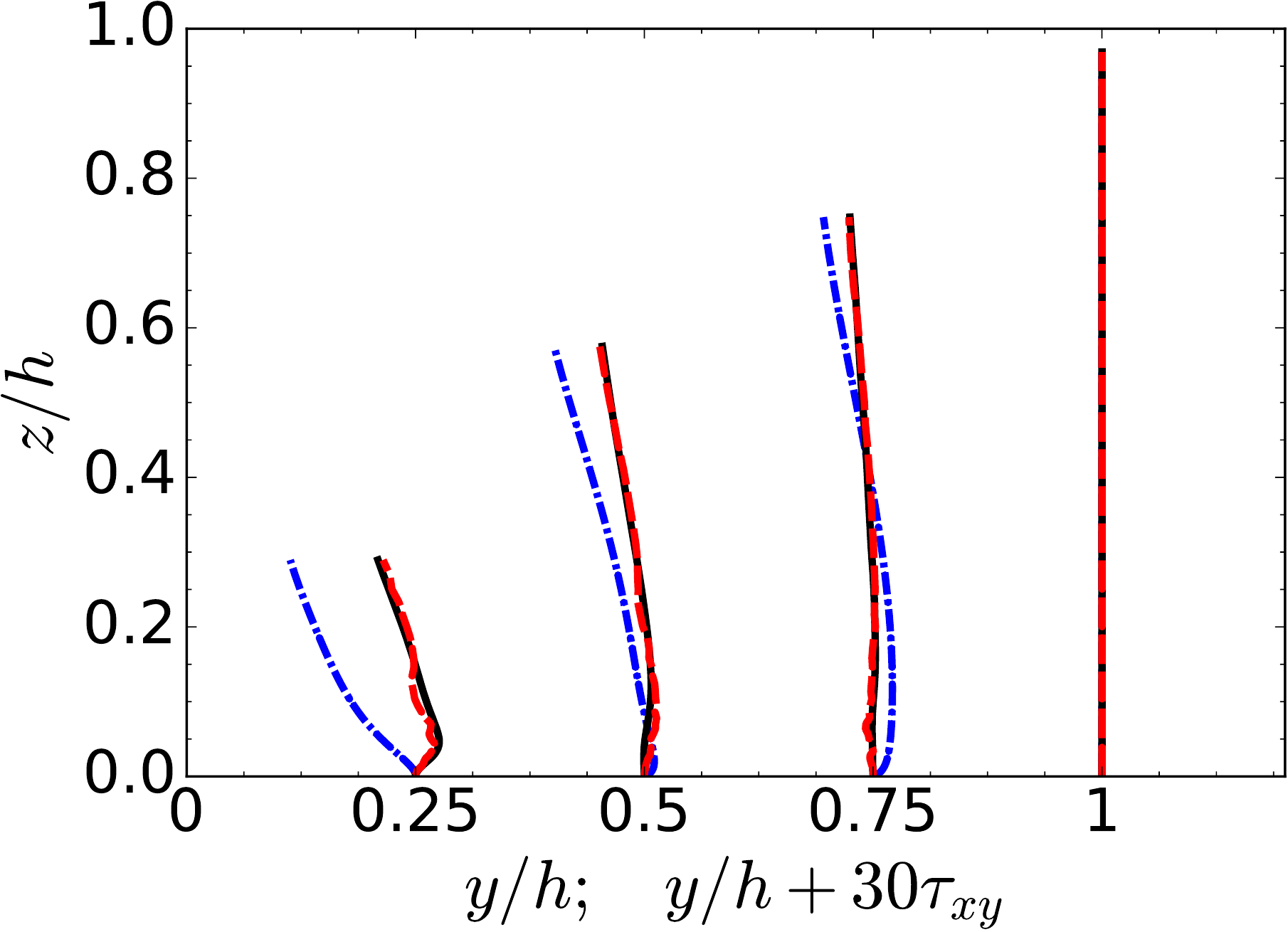}}\hspace{0.2em}
  \subfloat[$R_{xz}$]{\includegraphics[width=0.4\textwidth]{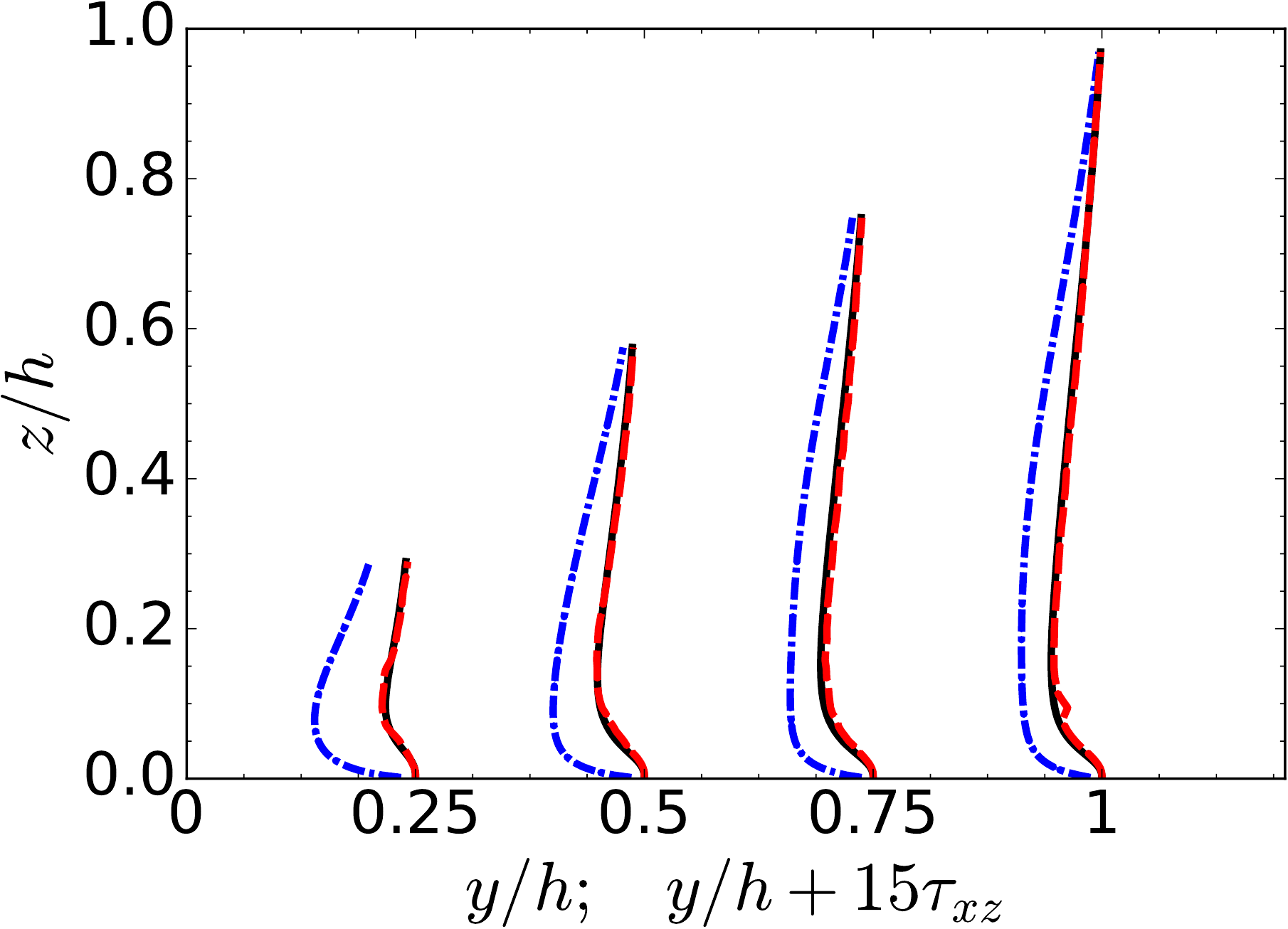}}
  \caption{The reconstructed Reynolds stress components based on the prediction of \textbf{unit quaternion} for the flow in a square duct at $Re=3500$, including (a) shear component $R_{xy}$ and (b) shear component $R_{xz}$. The training flow is the flow in a square duct at $Re=2900$.}
\label{fig:quaternion-Tau}
\end{figure}

The reconstructed Reynolds stress components $R_{xy}$ and $R_{xz}$ based on the prediction of unit quaternion also show satisfactory performance in Fig.~\ref{fig:quaternion-Tau}. Such satisfactory prediction performance is comparable to the reconstructed Reynolds stress based on the prediction of discrepancy-based Euler angles shown in Fig.~\ref{fig:euler-Tau}. It confirms that the similar prediction performance can be achieved for Reynolds stress components based on either discrepancy-based Euler angles or unit quaternion in this ideal training-prediction scenario.

\begin{figure}[!htbp]
  \centering
  {\includegraphics[width=0.3\textwidth]{component-legend}}\\
  \subfloat[$\Delta \phi_1$]{\includegraphics[width=0.32\textwidth]{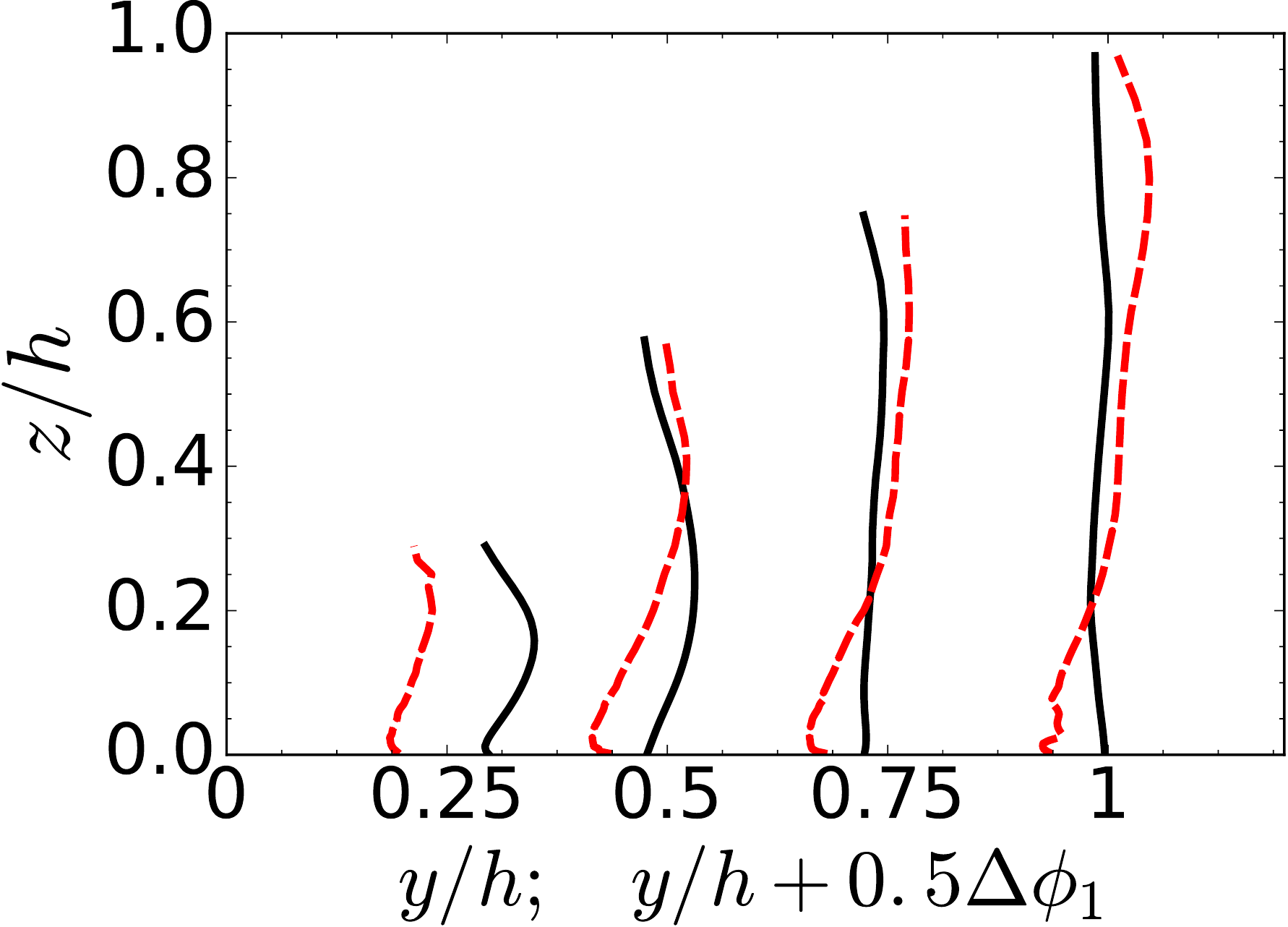}}\hspace{0.1em}
  \subfloat[$\Delta \phi_2$]{\includegraphics[width=0.32\textwidth]{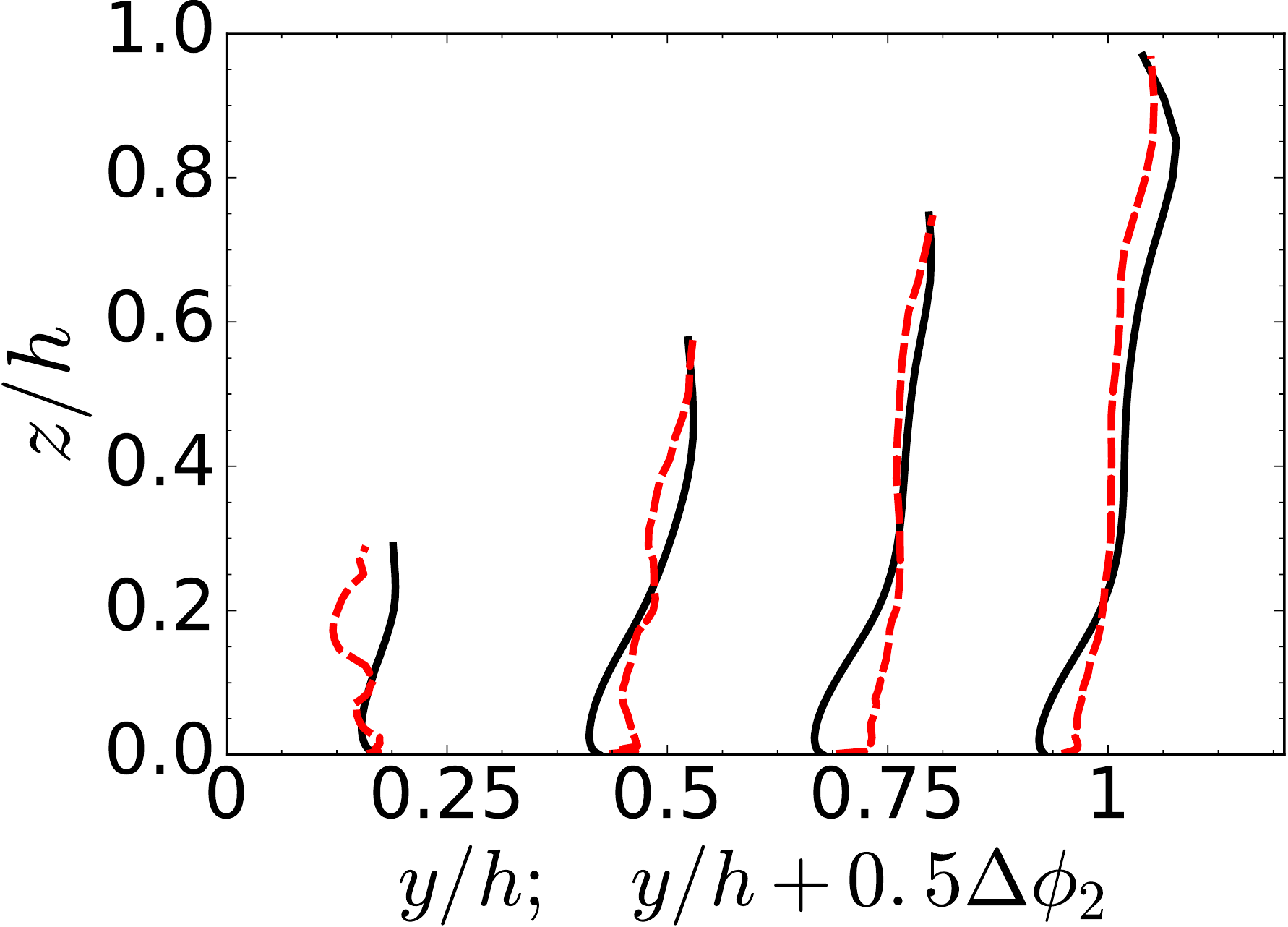}}\hspace{0.1em}
  \subfloat[$\Delta \phi_3$]{\includegraphics[width=0.32\textwidth]{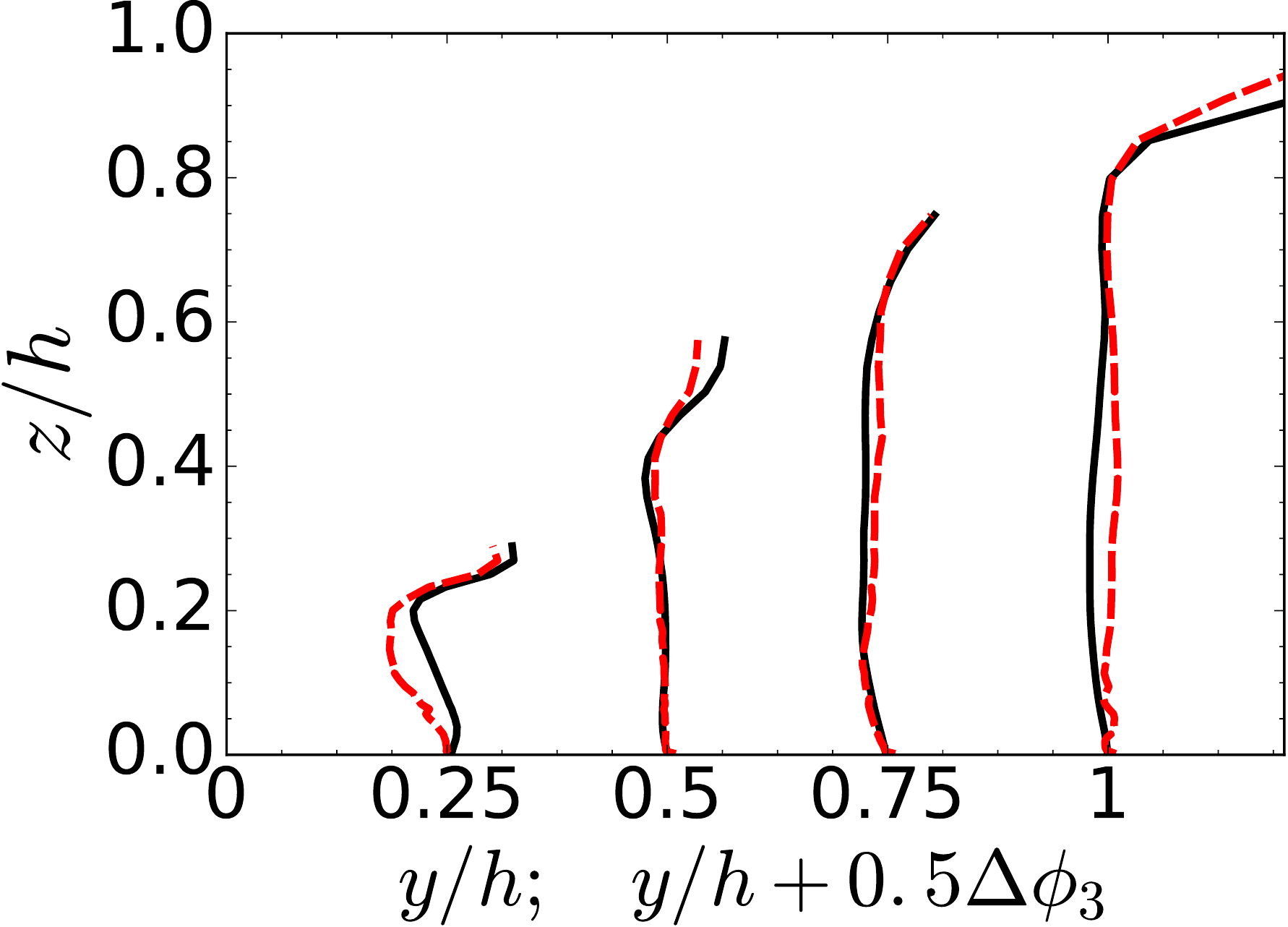}}
  \caption{The prediction of discrepancy-based Euler angles $\Delta \phi_\alpha$ for the flow in a square duct at $Re=3500$. The training flow is the flow in a square duct at $Re=2900$ with the coordinate system rotated by $60 \degree$.}
\label{fig:euler-rot}
\end{figure}

The second training-prediction scenario is chosen to demonstrate the shortcoming of the discrepancy-based Euler angles. The training flow and the test flow are the same as the first training-prediction scenario, while the coordinate system is rotated anti-clockwise by $60 \degree$ for the 
training flow. The coordinate system of the test flow remains unchanged. 
The objective of this training-prediction scenario is to mimic the possible situation in complex flows in industrial applications, where 
the training flow and the test flow locally share the similar flow physics but different flow direction or orientations. In this 
situation, it is unlikely that a choice of global coordinate system is able to take into account the difference between local 
flow directions, and thus the coordinate system relative to the flow direction would be different for the training 
flow and the test flow. The prediction of discrepancy-based Euler angles for the flow in a square duct at $Re=3500$ is shown in Fig.~\ref{fig:euler-rot}. Compared to the prediction performance as shown in Fig.~\ref{fig:euler}, the predicted discrepancy-based Euler angles in Fig.~\ref{fig:euler-rot} is less satisfactory when different coordinate systems are applied to the training flow and the test flow. Specifically, the deterioration of Euler angles is more pronounced for $\Delta \phi_1$ and $\Delta \phi_2$. This is because the discrepancy-based Euler angles are not invariants under the rotation of the reference frame.

\begin{figure}[!htbp]
  \centering
  {\includegraphics[width=0.3\textwidth]{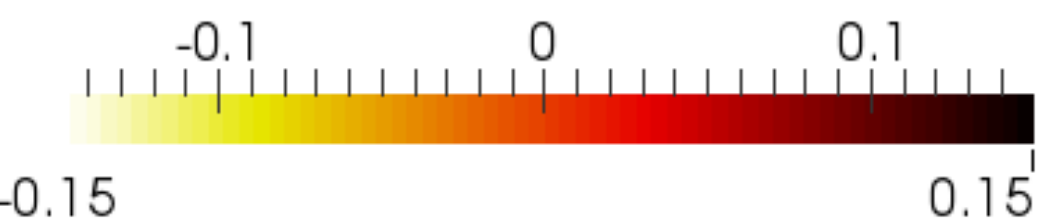}}\\
  \subfloat[Unrotated]{\includegraphics[width=0.3\textwidth]{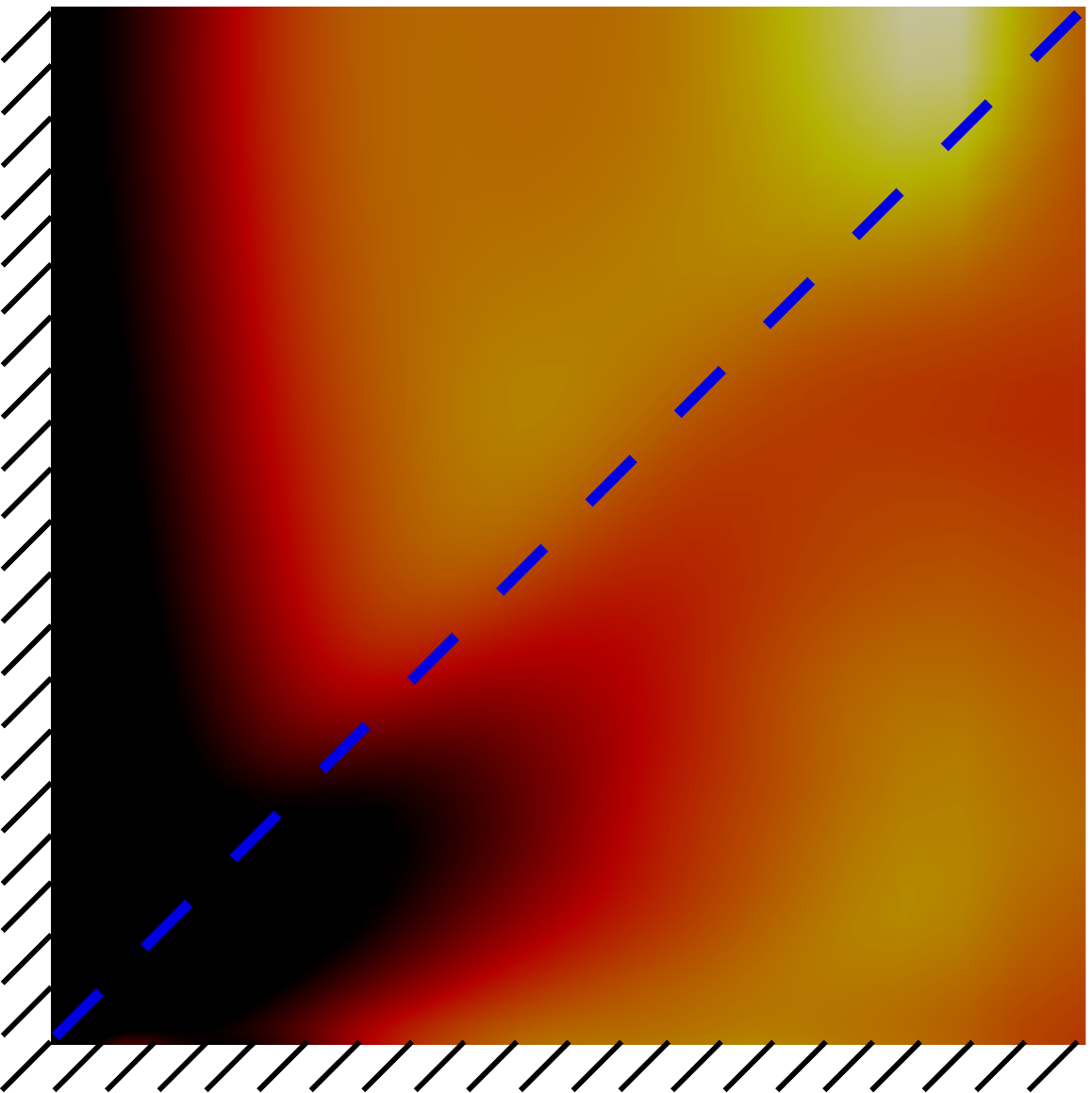}}\hspace{0.5em}
    \subfloat[Rotated]{\includegraphics[width=0.3\textwidth]{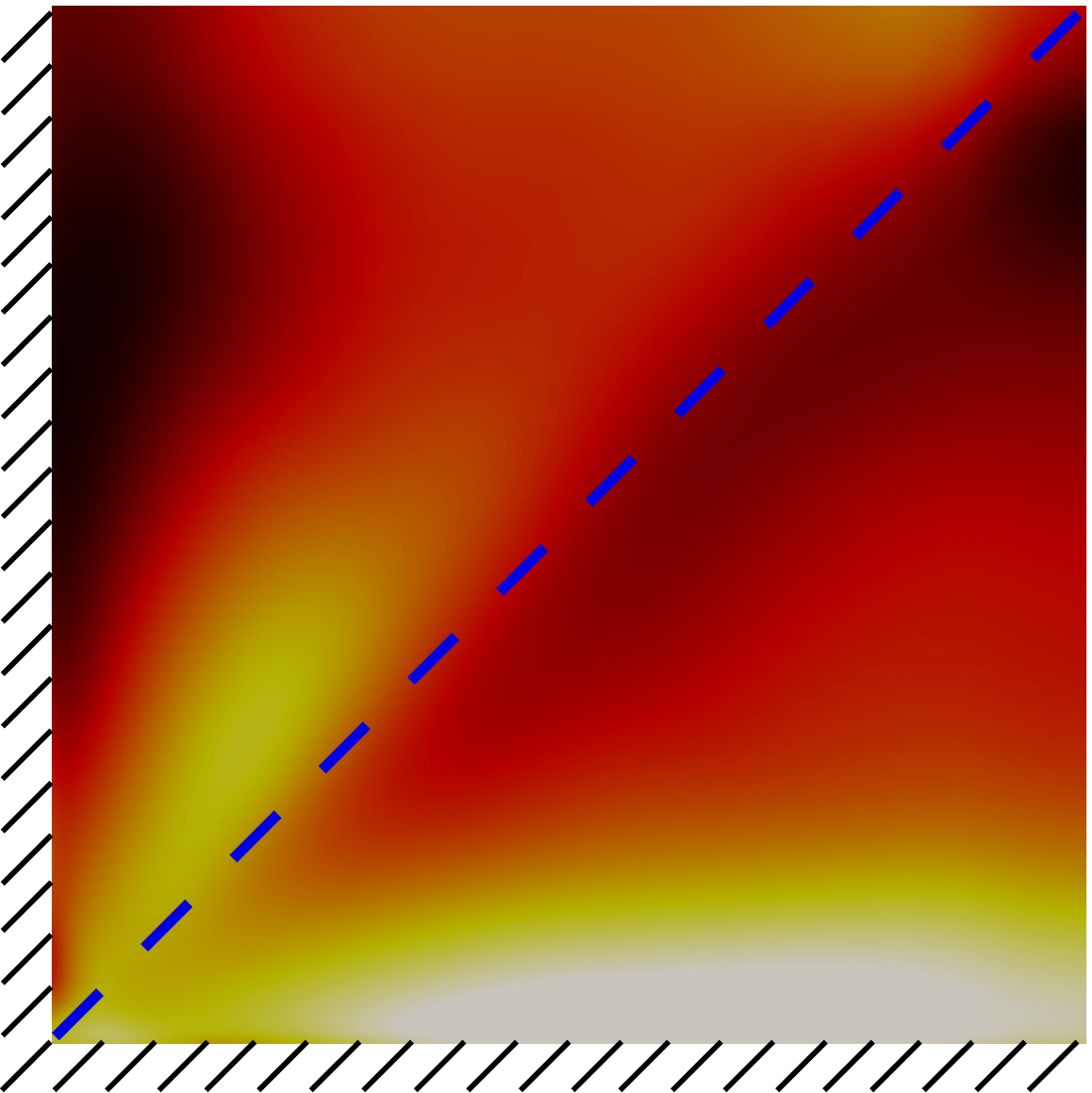}}\hspace{0.5em}
  \caption{The discrepancy-based Euler angles $\Delta \phi_1$ between the RANS simulated Reynolds stress and the DNS data of the training flow in a square duct at $Re=2900$. Panels (a) corresponds to the $\Delta \phi_1$ based on the original reference frame, for which the $y$ and $z$ axis align with the sidewall. Panels (b) corresponds to the $\Delta \phi_1$ based on the reference frame rotated by $60 \degree$ anti-clockwise. The dashed line denotes the diagonal of the square duct.}
\label{fig:euler-comp}
\end{figure}

The frame-dependence of discrepancy-based Euler angles is further demonstrated in Fig.~\ref{fig:euler-comp} by comparing the contours of discrepancy-based Euler angles based on the original reference frame and the rotated reference frame. It can be seen that discrepancy-based Euler angle $\Delta \phi_1$ changes with the rotation of the reference frame. Another component of the discrepancy-based Euler angles $\Delta \phi_2$ also changes with the rotation of the reference frame and is omitted here for brevity. The change of discrepancy-based Euler angles of training flow would lead to a different trained regression function, since the inputs mean flow features are all invariants under the rotation of reference frame~\cite{ling2015evaluation}. The frame-dependence nature of the discrepancy-based Euler angles explains the less satisfactory prediction performance as shown in Fig.~\ref{fig:euler-rot}. It demonstrates that the discrepancy-based Euler angles can potentially lead to poor prediction performance in the machine-learning-assisted turbulence modeling.

\begin{figure}[!htbp]
  \centering
  {\includegraphics[width=0.4\textwidth]{Tau-legend}}\\
  \subfloat[$R_{xy}$]{\includegraphics[width=0.4\textwidth]{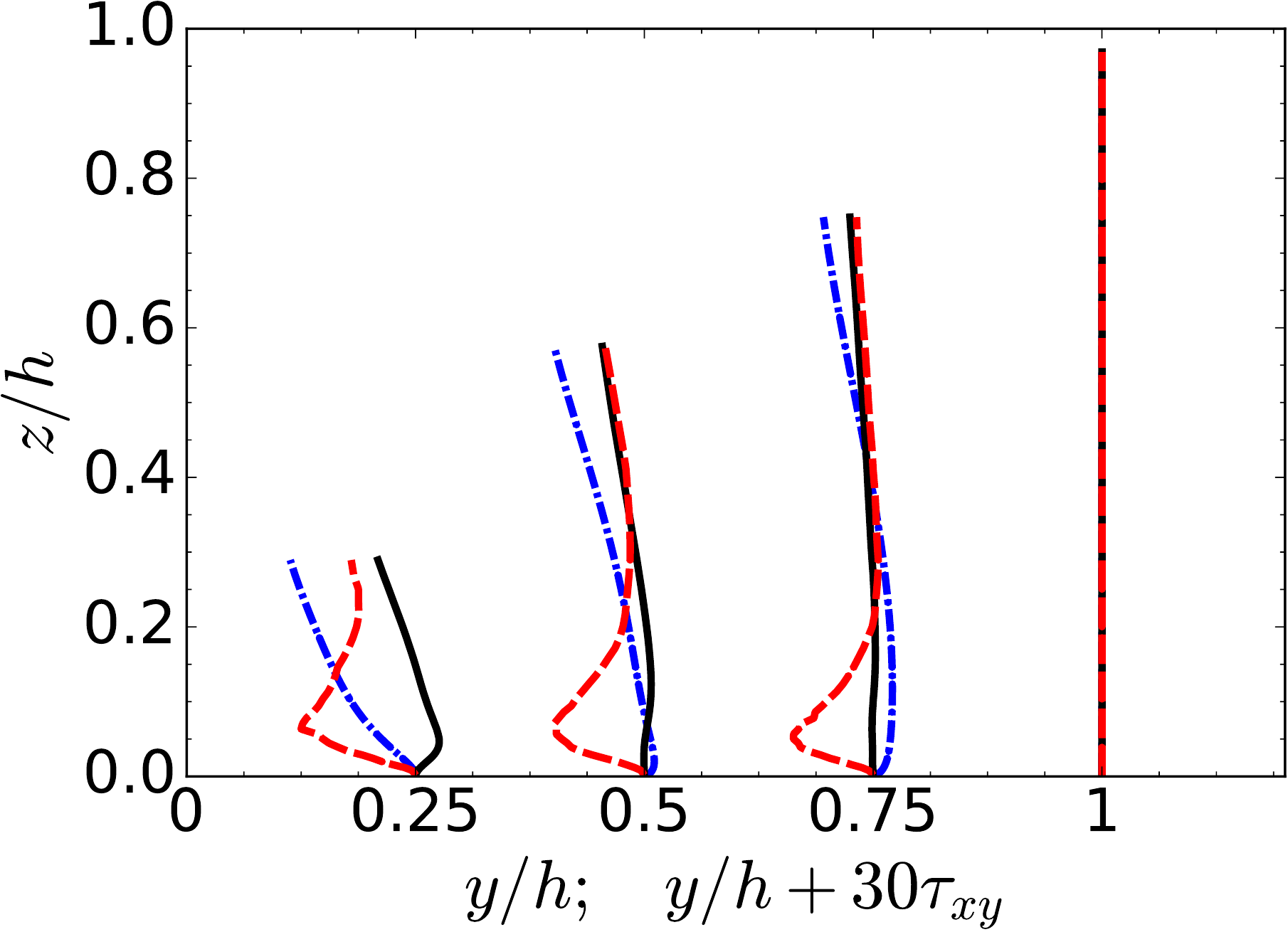}}\hspace{0.2em}
  \subfloat[$R_{xz}$]{\includegraphics[width=0.4\textwidth]{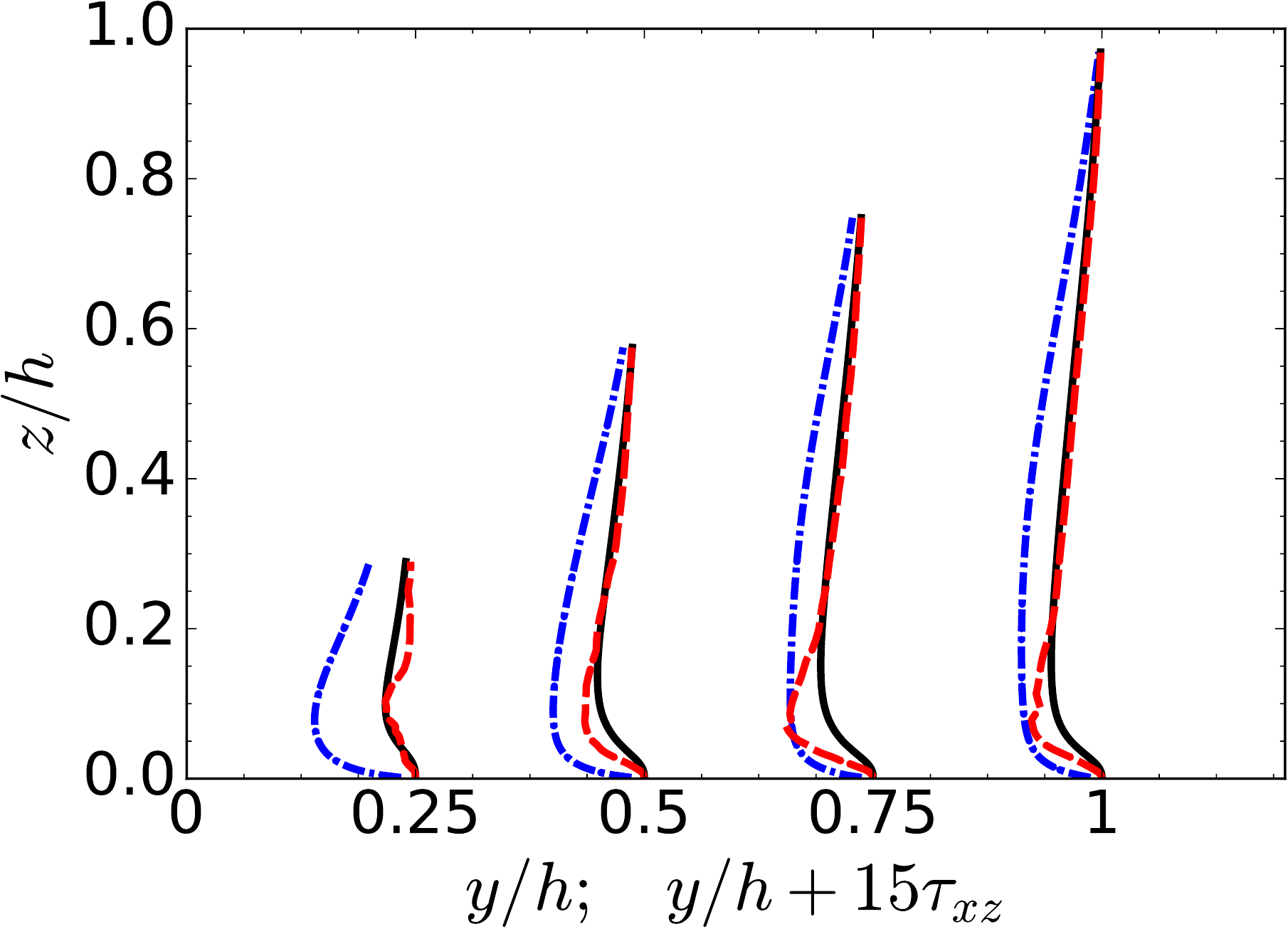}}
\caption{The prediction of Reynolds stress components based on $\textbf{discrepancy-based Euler angles}$ for the flow in a square duct at $Re=3500$. The training flow is the flow in a square duct at $Re=2900$ with the reference frame rotated by $60 \degree$ anti-clockwise.}
\label{fig:euler-rot-Tau}
\end{figure}

The reconstructed Reynolds stress components $R_{xy}$ and $R_{xz}$ based on the prediction of discrepancy-based Euler angles are shown in Fig.~\ref{fig:euler-rot-Tau} for the second training--prediction scenario. Compared to the reconstructed Reynolds stress as shown in Fig.~\ref{fig:euler-Tau}, the Reynolds stress components $R_{xy}$ and $R_{xz}$ show less satisfactory prediction performance. In the near-wall region, the prediction of the shear component $R_{xy}$ is even worse than the baseline RANS simulated results. It demonstrates that the reconstructed Reynolds stress via machine-learning-predicted discrepancy-based Euler angles is potentially unreliable, even for the scenario where training flow and test flow share similar flow physics and only differ by Reynolds number (a relatively easy case for machine learning).

\begin{figure}[!htbp]
  \centering
  {\includegraphics[width=0.4\textwidth]{Tau-legend}}\\
  \subfloat[Euler angles representation]{\includegraphics[width=0.6\textwidth]{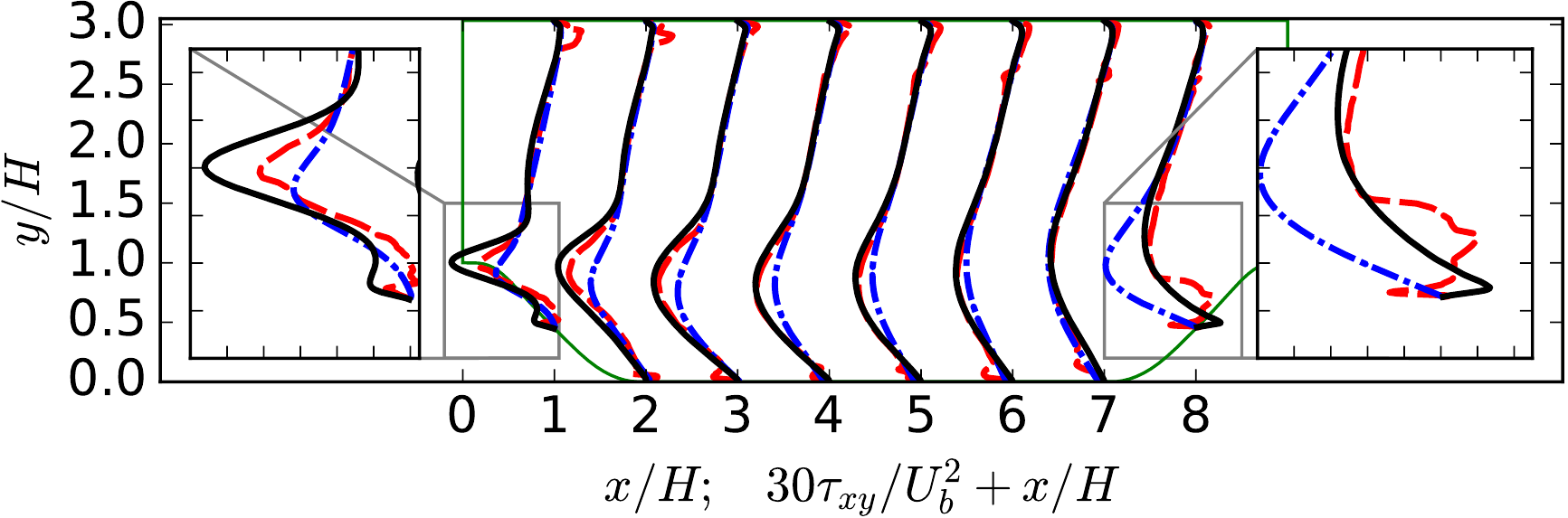}}\\
  \subfloat[Unit quaternion representation]{\includegraphics[width=0.6\textwidth]{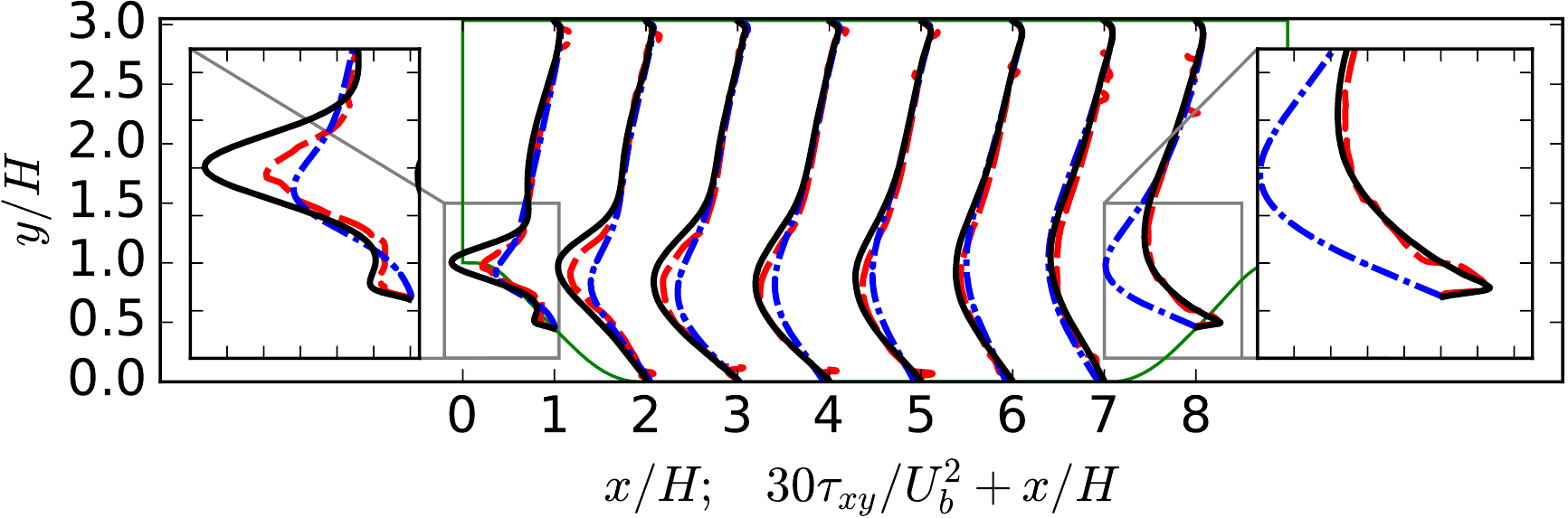}}\\
  \caption{The prediction of shear stress $R_{xy}$ based on (a) discrepancy-based Euler angles and (b) unit quaternion for the flow over periodic hills at $Re=5600$. The training flow is the flow over periodic hills at $Re=5600$ with a steeper hill profile. A zoom-in view is added at both the windward side and the leeward side of the hill for clear comparison.}
\label{fig:Tau-pehill}
\end{figure}

By investigating the third training-prediction scenario, we demonstrate the merit of unit quaternion representation in a more realistic application where training flow and test flow have different geometry configurations. The reconstructed Reynolds stress shear component $R_{xy}$ based on the prediction of discrepancy-based Euler angles is shown in Fig.~\ref{fig:Tau-pehill}a. It can be seen that the reconstructed shear stress component $R_{xy}$ is less satisfactory at the windward side of the hill, where the reconstructed $R_{xy}$ is even worse than the RANS results within the near wall region. The main reason is that the different steepness of hill profiles lead to different local flow directions for the training flow and the test flow at this region. Due to the frame-dependence nature of discrepancy-based Euler angles, the trained machine-learning function from the training flow is not applicable to the test flow. Therefore, the reconstructed Reynolds stress component based on the predicted discrepancy-based Euler angles of the test flow is potentially worse than the baseline RANS results. It should be noted that the deterioration of the reconstructed shear stress component $R_{xy}$ is less notable at the leeward of the hill, where the steepness of the hill profiles are also different for the training flow and the test flow. The main reason is that the turbulent kinetic energy is lower within the near wall region at leeward side of the hill, and thus the shifting of energy between the Reynolds stress components due to an inaccurate estimation of eigenvectors perturbation is less notable. Compared to the reconstructed Reynolds stress component in Fig.~\ref{fig:Tau-pehill}a, the reconstructed $R_{xy}$ based on the prediction of unit quaternion shows a much better agreement with the DNS data at the windward of the hill, as highlighted by a zoom-in view in Fig.~\ref{fig:Tau-pehill}. For the leeward side of the hill, the reconstructed $R_{xy}$ in Fig.~\ref{fig:Tau-pehill}b is also slightly better than the one shown in Fig.~\ref{fig:Tau-pehill}a. Therefore, unit quaternion provides a better representation of the eigenvectors perturbations to the Reynolds stress tensor in the context of machine-learning-assisted turbulence modeling.

\section{Conclusion}
\label{sec:conclusion}

Introducing perturbations to stress tensors has important implications to model-form uncertainty
estimation in RANS models and to machine-learning-assisted RANS modeling.  Representing perturbations
to Reynolds stress eigenvectors in the context of machine-learning-assisted modeling is challenging
due to the requirements of mapping smoothness and frame-independence.  In this work we formulated
the eigenvector perturbations as rigid-body rotations and examined three representations: (1) the
direct-rotation-based Euler angles, (2) the discrepancy-based Euler angles, and (3) the unit quaternion. A
priori assessment of fully turbulent square duct flow shows that the direct-rotation-based Euler
angles representation does not satisfy the smoothness requirement. On the other hand, the
discrepancy-based Euler angle representation does not satisfy the frame-independence requirement, which
has been shown theoretically and numerically with \textit{a posteriori} tests on both the square duct flow
and the flow over periodic hills. The numerical examples showed that the unit quaternion
representation satisfies both requirements and is an ideal representation of Reynolds stress eigenvector
perturbations. This finding has clear importance for uncertainty quantification and machine learning
in turbulence modeling and for data-driven computational mechanics in general. Moreover, it also has
implications in other fields such as plasticity, where sequential increments of stress tensors are
used to find the new stress state from the current one.

\section{Acknowledgment}
The authors would like to acknowledge Dr. Roney L. Thompson of Federal University of Rio de Janeiro for providing us helpful discussions during this work.





\bibliographystyle{elsarticle-num}

\begin{thebibliography}{10}
\expandafter\ifx\csname url\endcsname\relax
  \def\url#1{\texttt{#1}}\fi
\expandafter\ifx\csname urlprefix\endcsname\relax\def\urlprefix{URL }\fi
\expandafter\ifx\csname href\endcsname\relax
  \def\href#1#2{#2} \def\path#1{#1}\fi

\bibitem{wilcox1998turbulence}
D.~C. Wilcox, Turbulence modeling for CFD, 3rd Edition, DCW Industries, 2006.

\bibitem{wang2017physics}
J.-X. Wang, J.-L. Wu, H.~Xiao, Physics-informed machine learning approach for
  reconstructing reynolds stress modeling discrepancies based on {DNS} data,
  Physical Review Fluids 2~(3) (2017) 034603.

\bibitem{ling2016reynolds}
J.~Ling, A.~Kurzawski, J.~Templeton, Reynolds averaged turbulence modelling
  using deep neural networks with embedded invariance, Journal of Fluid
  Mechanics 807 (2016) 155--166.

\bibitem{singh2016machine}
A.~P. Singh, S.~Medida, K.~Duraisamy, Machine learning-augmented predictive
  modeling of turbulent separated flows over airfoils, arXiv preprint
  arXiv:1608.03990.

\bibitem{tracey2015machine}
B.~Tracey, K.~Duraisamy, J.~J. Alonso, A machine learning strategy to assist
  turbulence model development, AIAA Paper 1287 (2015) 2015.

\bibitem{singh16using}
A.~P. Singh, K.~Duraisamy, Using field inversion to quantify functional errors
  in turbulence closures, Physics of Fluids 28 (2016) 045110.

\bibitem{popebook}
S.~B. Pope, Turbulent Flows, Cambridge University Press, Cambridge, 2000.

\bibitem{xiao2017perspectives}
H.~Xiao, W.~J.-L., W.~J.-X., E.~Paterson, Physics-informed machine learning for
  predictive turbulence modeling: Progress and perspectives, AIAA SciTech
  Meeting (2017) 2017--1712\href {http://dx.doi.org/AIAA-2017-1712}
  {\path{doi:AIAA-2017-1712}}.

\bibitem{king2016autonomic}
R.~N. King, P.~E. Hamlington, W.~J. Dahm, Autonomic closure for turbulence
  simulations, Physical Review E 93~(3) (2016) 031301.

\bibitem{vollant2017subgrid}
A.~Vollant, G.~Balarac, C.~Corre, Subgrid-scale scalar flux modelling based on
  optimal estimation theory and machine-learning procedures, Journal of
  Turbulence (2017) 1--25.

\bibitem{emory2013modeling}
M.~Emory, J.~Larsson, G.~Iaccarino, Modeling of structural uncertainties in
  {Reynolds}-averaged {Navier}-{Stokes} closures, Physics of Fluids 25~(11)
  (2013) 110822.

\bibitem{emory2011modeling}
M.~Emory, R.~Pecnik, G.~Iaccarino, Modeling structural uncertainties in
  {Reynolds-averaged} computations of shock/boundary layer interactions, AIAA
  paper 479 (2011) 1--16.

\bibitem{emory14estimate}
M.~A. Emory, Estimating model-form uncertainty in {Reynolds-averaged}
  {Navier--Stokes} closures, Ph.D. thesis, Stanford University (2014).

\bibitem{banerjee2007presentation}
S.~Banerjee, R.~Krahl, F.~Durst, C.~Zenger, Presentation of anisotropy
  properties of turbulence, invariants versus eigenvalue approaches, Journal of
  Turbulence 8~(32) (2007) 1--27.

\bibitem{lumley1977return}
J.~L. Lumley, G.~R. Newman, The return to isotropy of homogeneous turbulence,
  Journal of Fluid Mechanics 82~(01) (1977) 161--178.

\bibitem{roney16strategy}
R.~Thompson, L.~Sampaio, W.~Edeling, A.~A. Mishra, G.~Iaccarino, A strategy for
  the eigenvector perturbations of the reynolds stress tensor in the context of
  uncertainty quanti?cation, in: Proceedings of the Summer Program, Center
  for Turbulence Research, 2016, p.~10.

\bibitem{iaccarino2017eigenspace}
G.~Iaccarino, A.~A. Mishra, S.~Ghili, Eigenspace perturbations for uncertainty
  estimation of single-point turbulence closures, Physical Review Fluids 2~(2)
  (2017) 024605.

\bibitem{mfu5}
J.-X. Wang, R.~Sun, H.~Xiao, Quantification of uncertainties in turbulence
  modeling: A comparison of physics-based and random matrix theoretic
  approaches, International Journal of Heat and Fluid Flows 62 (2016) 577--592.

\bibitem{wu2017priori}
J.-L. Wu, J.-X. Wang, H.~Xiao, J.~Ling, A priori assessment of prediction
  confidence for data-driven turbulence modeling, Flow, Turbulence and
  CombustionIn press, doi: 10.1007/s10494-017-9807-0.

\bibitem{huynh2009metrics}
D.~Q. Huynh, Metrics for 3d rotations: Comparison and analysis, Journal of
  Mathematical Imaging and Vision 35~(2) (2009) 155--164.

\bibitem{kuffner2004effective}
J.~J. Kuffner, Effective sampling and distance metrics for 3d rigid body path
  planning, in: Robotics and Automation, 2004. Proceedings. ICRA'04. 2004 IEEE
  International Conference on, Vol.~4, IEEE, 2004, pp. 3993--3998.

\bibitem{horn1987closed}
B.~K. Horn, Closed-form solution of absolute orientation using unit
  quaternions, JOSA A 4~(4) (1987) 629--642.

\bibitem{heeger1990simple}
D.~J. Heeger, A.~Jepson, Simple method for computing 3d motion and depth, in:
  Computer Vision, 1990. Proceedings, Third International Conference on, IEEE,
  1990, pp. 96--100.

\bibitem{domingos2012few}
P.~Domingos, A few useful things to know about machine learning, Communications
  of the ACM 55~(10) (2012) 78--87.

\bibitem{oliver2011bayesian}
T.~A. Oliver, R.~D. Moser, Bayesian uncertainty quantification applied to
  {RANS} turbulence models, in: Journal of Physics: Conference Series, Vol.
  318, IOP Publishing, 2011, p. 042032.

\bibitem{wang2017predictive}
J.-X. Wang, J.-L. Wu, J.~Ling, G.~Iaccarino, H.~Xiao, Physics-informed machine
  learning for predictive turbulence modeling: Towards a complete framework,
  Tech. rep., Center of Turbulence Research, Proceedings of the Summer Program,
  Stanford University (2016).

\bibitem{simonsen2005turbulent}
A.~Simonsen, P.-{\AA}. Krogstad, Turbulent stress invariant analysis:
  Clarification of existing terminology, Physics of Fluids 17~(8) (2005)
  088103.

\bibitem{pope1975general}
S.~Pope, A more general effective-viscosity hypothesis, Journal of Fluid
  Mechanics 72~(02) (1975) 331--340.

\bibitem{ling2016JCP}
J.~Ling, R.~Jones, J.~Templeton, Machine learning strategies for systems with
  invariance properties, Journal of Computational Physics 318 (2016) 22--35.

\bibitem{goldstein1980euler}
H.~Goldstein, The {Euler} angles, Classical Mechanics, (1980) 143--148.

\bibitem{kuipers1999quaternions}
J.~B. Kuipers, et~al., Quaternions and rotation sequences, Vol.~66, Princeton
  University press Princeton, 1999.

\bibitem{huser1993direct}
A.~Huser, S.~Biringen, Direct numerical simulation of turbulent flow in a
  square duct, Journal of Fluid Mechanics 257 (1993) 65--95.

\bibitem{weller1998tensorial}
H.~G. Weller, G.~Tabor, H.~Jasak, C.~Fureby, A tensorial approach to
  computational continuum mechanics usi ng ject-oriented techniques, Computers
  in physics 12~(6) (1998) 620--631.

\bibitem{patankar1980numerical}
S.~Patankar, Numerical heat transfer and fluid flow, CRC press, 1980.

\bibitem{gibson1978ground}
M.~Gibson, B.~Launder, Ground effects on pressure fluctuations in the
  atmospheric boundary layer, Journal of Fluid Mechanics 86~(03) (1978)
  491--511.

\bibitem{pinelli2010reynolds}
A.~Pinelli, M.~Uhlmann, A.~Sekimoto, G.~Kawahara, Reynolds number dependence of
  mean flow structure in square duct turbulence, Journal of Fluid Mechanics 644
  (2010) 107--122.

\bibitem{laizet2011incompact3d}
S.~Laizet, N.~Li, Incompact3d: A powerful tool to tackle turbulence problems
  with up to o (105) computational cores, International Journal for Numerical
  Methods in Fluids 67~(11) (2011) 1735--1757.

\bibitem{laizet2009high}
S.~Laizet, E.~Lamballais, High-order compact schemes for incompressible flows:
  A simple and efficient method with quasi-spectral accuracy, Journal of
  Computational Physics 228~(16) (2009) 5989--6015.

\bibitem{breuer2008}
M.~Breuer, N.~Peller, C.~Rapp, M.~Manhart, Flow over periodic hills: Numerical
  and experimental study in a wide range of reynolds numbers, Computers \&
  Fluids 38~(2) (2009) 433--457.

\bibitem{launder1974application}
B.~Launder, B.~Sharma, Application of the energy-dissipation model of
  turbulence to the calculation of flow near a spinning disc, Letters in Heat
  and Mass Transfer 1~(2) (1974) 131--137.

\bibitem{breuer2009flow}
M.~Breuer, N.~Peller, C.~Rapp, M.~Manhart, Flow over periodic hills--numerical
  and experimental study in a wide range of reynolds numbers, Computers \&
  Fluids 38~(2) (2009) 433--457.

\bibitem{ling2015evaluation}
J.~Ling, J.~Templeton, Evaluation of machine learning algorithms for prediction
  of regions of high {Reynolds averaged Navier Stokes} uncertainty, Physics of
  Fluids (1994-present) 27~(8) (2015) 085103.

\end{thebibliography}

\appendix

\section*{Nomenclature}

\begin{tabbing}
  XXXX \= \kill
  $\Delta$ \> discrepancy \\
  $\mathbf{\Lambda}$ \>  diagonal matrix of eigenvalues \\
  $\mathbf{\Omega}$ \> rotation rate tensor \\
  $\gamma$ \> ratio of turbulent kinetic energy \\
  $\mathbf{R}$ \> Reynolds stress \\
  $\phi_{\alpha}$ \> global-frame-based Euler angles \\
  $\phi_{\alpha}^{o \to *}$ \> direct-rotation-based Euler angles \\
  $\mathbf{S}$ \> strain rate tensor \\
  $\mathbf{h}$ \> unit quaternion \\
  $\mathbf{q}$ \> mean flow features \\
  $\mathbf{n}$ \> Euler axis \\
  $\mathbf{Q}$ \>  rotation matrix \\
  $\mathbf{V}$ \>  eigenvectors of second order tensor \\
  $\nu_t$ \> turbulent viscosity
   \end{tabbing}
\end{document}